\documentclass[acmtog]{acmart}
\pdfoutput=1

\usepackage{booktabs} 

\setcitestyle{square}

\usepackage[linesnumbered,ruled,vlined]{algorithm2e} 

\SetAlFnt{\small}
\SetAlCapFnt{\small}
\SetAlCapNameFnt{\small}
\SetAlCapHSkip{0pt}
\IncMargin{-\parindent}

\usepackage{tikz}
\usepackage{wrapfig}
\usepackage{subcaption}
\usepackage{microtype}
\usepackage{multirow}

\newcommand{\eg}{\emph{e.g.}\xspace}
\newcommand{\ie}{\emph{i.e.}\xspace}
\newcommand{\cf}{\emph{cf.}\xspace}

\setcopyright{none}

\begin{document}
\title{Layered Fields for Natural Tessellations on Surfaces}

\author{Rhaleb Zayer }
\affiliation{%
	\institution{Max Planck Institute for Informatics}
	\institution{Saarland Informatics Campus, Germany}
}
\email{rzayer@mpi-inf.mpg.de}

\author{Daniel Mlakar}
\affiliation{%
	\institution{Graz University of Technology}
	\institution{Graz, Austria}
}
\email{daniel.mlakar@icg.tugraz.at}

\author{Markus Steinberger}
\affiliation{%
	\institution{Graz University of Technology}
	\institution{Graz, Austria}
}
\email{steinberger@icg.tugraz.at}

\author{Hans-Peter Seidel}
\affiliation{%
	\institution{Max Planck Institute for Informatics}
	\institution{Saarland Informatics Campus, Germany}
}
\email{hpseidel@mpi-inf.mpg.de}

\renewcommand{\shortauthors}{Rhaleb Zayer, Daniel Mlakar, Markus Steinberger, Hans-Peter Seidel}

\begin{abstract}
Mimicking natural tessellation patterns is a fascinating multi-disciplinary problem. Geometric methods aiming at reproducing such partitions on surface meshes are commonly based on the Voronoi model and its variants, and are often faced with challenging issues such as metric estimation, geometric, topological complications, and most critically parallelization. In this paper, we introduce an alternate model which may be of value for resolving these issues.
We drop the assumption that regions need to be separated by lines. Instead, we regard region boundaries as narrow bands and we model the partition as a set of smooth functions layered over the surface. Given an initial set of seeds or regions, the partition emerges as the solution of a time dependent set of partial differential equations describing concurrently evolving fronts on the surface. Our solution does not require geodesic estimation, elaborate numerical solvers, or complicated bookkeeping data structures. The cost per time-iteration is dominated by the multiplication and addition of two sparse matrices. Extension of our approach in a Lloyd's algorithm fashion can be easily achieved and the extraction of the dual mesh can be conveniently preformed in parallel through matrix algebra. As our approach relies mainly on basic linear algebra kernels, it lends itself to efficient implementation on modern graphics hardware.

\end{abstract}

\maketitle

\section{Introduction}
\label{sec:intro}

\begin{figure}[t]
  \centering
  \includegraphics[width=1\linewidth]{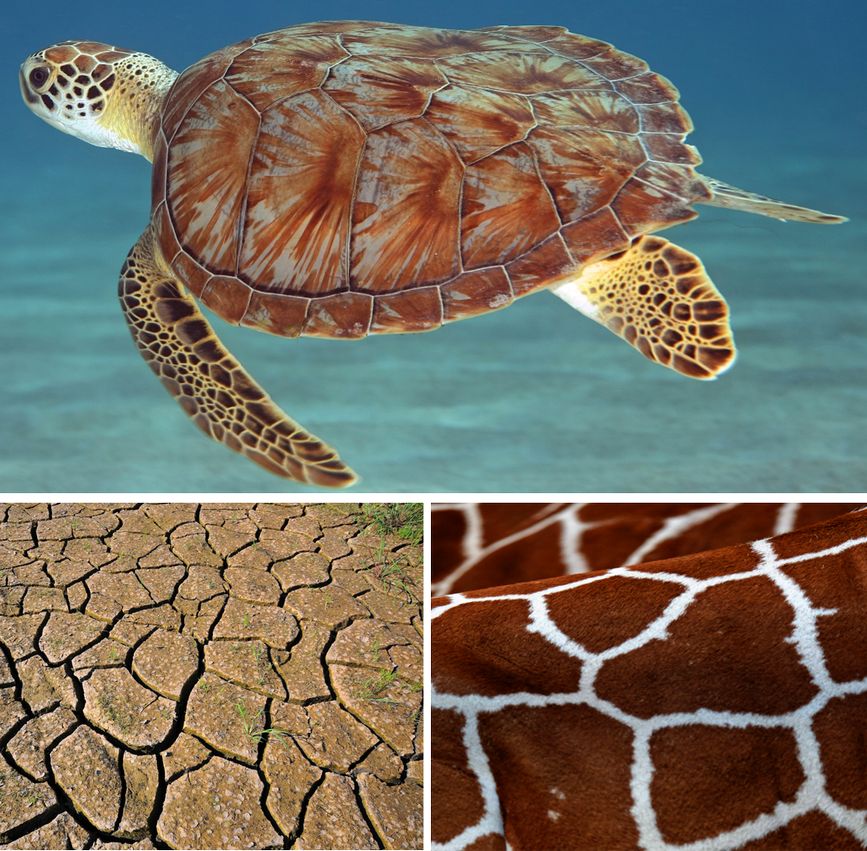}
  \caption{Natural tessellations as seen on a tortoise shell (top), drying earth (bottom-left), and a giraffe pattern. Cells are are not separated by razor thin lines but rather by narrow bands which are not necessarily straight.
  }
  \label{fig:natureImages}
\end{figure}

In his work on celestial bodies, Descartes devised diagrams where planar points (generators, sites) are separated by lines equidistant to their closest neighboring points~\cite{Okabe:1992:STC}. The resulting regions enclosed by the lines are called Voronoi cells in honor of Voronoi who studied the problem extension to higher dimensions extensively.
A restriction which imposes that generators coincide with the center of mass of their respective cells, yields the so called centroidal Voronoi tessellations (CVT). In general, CVT leads to more balanced cells and point distribution as the number of generators increases (Gersho's conjuncture~\cite{Gersho:1979:AOBQ}). In the planar and volumetric settings, these methods have been used across multiple disciplines ranging from geographic planing to architectural design. Their aesthetic appeal arises from invoking the beauty of natural tessellations, see Figure~\ref{fig:natureImages}. The direct creation of these tessellations on surfaces however, is inhibited by difficulties arising from estimating metrics on curved surfaces and by the toil of maintaining a valid secondary mesh, geometrically and topologically, which evolves on top of the original one.

In this work, we re-examine the notion of Voronoi diagrams and propose an alternate model which can simplify the creation of natural tessellations on surface meshes.
We regard Voronoi diagrams as an idealization of observable natural phenomena. In this idealization, boundaries across regions are modeled as razor sharp lines. In nature, however, this is seldom the case as shown in Figure~\ref{fig:natureImages}. Furthermore, the natural mechanisms which steer the creation of these cell-like structures often do not require knowledge of the location of other seeds or the geodesic paths towards them. Different fronts grow simultaneously till they collide and form cell walls. In order to capture this behavior we rely on an idea introduced by Fix to address free boundary problems, and in particular the Stefan type~\cite{Fix:1981:PFM}. In the numerical modeling of such problems, \eg, melting ice, tracking the location of the sharp interface between the solid and liquid states over time complicates the problem setup. On the other hand, treating the interface as very small band where the transition from one state to another is continuous leads to a much more tractable problem formulation.

We drop the assumption that tessellation cells need to be separated by lines. Instead, we regard cell boundaries as narrow bands and we model the partition as a set of smooth functions layered over the surface. Given an initial set of seeds or regions, the partition is obtained as the solution of a time dependent set of partial differential equations describing concurrently evolving fronts. The numerical solution does not require elaborate solvers and is carried out by a simple Euler time-stepping scheme. Furthermore, we show how the problem can be formulated for efficient implementation. In this way, elaborate mesh data structures can be avoided in favor of sparse matrix algebra. Within this formalism, the cost per time-iteration is dominated by the multiplication of two sparse matrices. The extension of our approach in a Lloyd's algorithm fashion~\cite{Lloyd:1982:LSQ} is straightforward. As a proof of concept, using an extremely simple approximation of cell centroids, our approach produces well balanced partitions. Our formulation is intended for producing partitions separated by smooth narrow bands. Nonetheless, sharp boundaries can be obtained at will. Furthermore, our approach allows generating and monitoring the quality of the Delaunay-like dual mesh throughout Lloyd's iterations without compromising performance. As our approach relies mainly on basic linear algebra kernels, it lends itself to efficient implementation on modern graphics hardware and we provide the necessary steps for reproducing our performance.

\section{Related work}
\label{related}

\paragraph{Historic developments:}
CVT can be achieved by the probabilistic method proposed by Macqueen~\cite{macqueen1967} which performs generator updates by taking the mean of a randomly sampled point (Monte-carlo method) and its closest generator. A parallel implementation and several improvements of this approach have been proposed by \citeauthor{Ju:2002:PMCVT}~\shortcite{Ju:2002:PMCVT}. Alternatively, an initial Voronoi diagram can be iteratively updated by taking the current configuration centroids as the generators for the next Voronoi diagram~\cite{Lloyd:1982:LSQ}. Reformulation of the problem as an objective function which can be addressed by means of Newton-like approaches has been studied by \citeauthor{Iri:1984}~\shortcite{Iri:1984} as well as \citeauthor{Du:2006:ASCCVT}~\shortcite{Du:2006:ASCCVT}.
\paragraph{Sampling and partitioning of meshes}
The above methods apply to both planar and volumetric settings as Euclidean distance metrics can be directly used and are easy to compute. The extension to surfaces is impeded by the need to measure or approximate metrics on surfaces. This limitation poses a great challenge in computer graphics and steered a steady research effort dedicated to addressing various facets of the problem. Turk~\shortcite{Turk:1992:RPS} developed a variant of Lloyd's algorithm which proceeds from a random seed configuration on the surfaces and uses repelling forces to redistribute the seeds evenly. The method requires a local planar unfolding to estimate the repulsion force. By working on implicit surfaces, local flattening can be avoided~\cite{Witkin:1994}. The final configuration marks the equilibrium of a set of constrained control points and freely floating particles which can merge or vanish driven by potentials modeled as Gaussian kernels.

Although the CVT problem can be addressed by means of restricted Voronoi diagrams, \ie, the intersection of the mesh and a volumetric Voronoi diagram.
This approach requires extensive data structure management, may not scale well for large meshes, and faces additional challenges resulting from complicated geometric or topological figures. We do not cover these methods in this brief overview and restrict the discussion to methods which operate directly on surface meshes. In the work of \citeauthor{Alliez:2003:ISR}~\shortcite{Alliez:2003:ISR}, the metric problem was avoided by directly flattening the surface and performing Lloyd iterations in the parametric plane. Instead of a global parametrization, the approach of \citeauthor{SurazhskyAlliezGotsman2003} \shortcite{SurazhskyAlliezGotsman2003} uses a patch based parametrization to perform Lloyd's updates on the local parametric domain. The approach of~\citeauthor{Peyre:2006} \shortcite{Peyre:2006} performs CVT on surfaces by taking advantage of the fast marching approach estimation of geodesics~\cite{Kimmel:98:CGPM} which extends the level set approach for solving the eikonal equation~\cite{Sethian:1996} to triangle meshes. The authors further use a gradient descent for finding local intrinsic centers of mass.

In a follow-up to Newton like approaches~\cite{Iri:1984,Du:2006:ASCCVT}, it has been shown that an L-BFGS formulation of CVT is in general faster than the standard Lloyd approach~\cite{Liu:2009:CVT}. The estimation of CVT on surfaces however is performed by means of standard Euclidean metrics similar to \citeauthor{Du:2003:CCVTS} \shortcite{Du:2003:CCVTS} and requires careful face triangle splitting along the boundaries of Voronoi cells. The approach of \citeauthor{Xin:2016:CPD} \shortcite{Xin:2016:CPD} extends the latter to power diagrams, and allow generators to evolve on the tangent plane and re-project them on the surface for evaluation of the objective function. The extensive evaluations of geodesics between points~\cite{Surazhsky:2005:FEAGM}, however restricts the method to very small meshes ($1K$ triangles) and limits its practical scope. Although, performance can be possibly improved by using more recent geodesic approximations~\cite{Crane:2013:GHN,Ying:2013:SVG}, it is not expected to drop significantly due to the high number of required queries. The work of~\citeauthor{Wang:2015:ICCVT} \shortcite{Wang:2015:ICCVT} proposes several enhancements to the standard LLoyd algorithm by means of local exponential maps and also extends the same ideas to the framework of \citeauthor{Liu:2009:CVT} \shortcite{Liu:2009:CVT}.

With a few exceptions, most of existing work avoids direct evaluations on meshes by some sort of global or local planar embedding. The need to compute geodesics, often done by third party code, maintain an evolving mesh topology, and perform several non trivial geometric and topological operations, makes the implementation of these methods a daunting task. The resulting not-so-straightforward algorithmic pipeline complicates the portability of these methods to modern parallel computing hardware.
From a simulation standpoint, early methods for producing patterns on surfaces, \eg, \citeauthor{Turk:1991:GTA} \shortcite{Turk:1991:GTA} can yield pleasant Voronoi-like patterns, however the controllability of these approaches and the difficulties in extracting relevant cells inhibits the adoption of these methods as geometric partitioning tools.

Most recently, \citeauthor{Herholz:dd:2017} \shortcite{Herholz:dd:2017} take advantage of the heat diffusion approach~\cite{Crane:2013:GHN} for approximating geodesic distances to drive the tessellation generation.
Although they parallelize their approach on the GPU, they suffer from major scalability issues. Besides the computational and memory costs induced by factorization, they need to determine region affiliation, which requires comparison between the heat diffusion solutions, \eg, for 10 thousand seeds and a mesh of 10 million vertices, they require cross row comparison in a $10k \times 10M$ matrix to find the correct approximation of Voronoi Cells.
To reduce comparisons, they estimate a radius for each seed by executing Dijkstra's algorithm, which by itself has similar complexity to the overall heat diffusion; they ignore the cost of this step in their timings.
Furthermore, this estimate fails for non-regular seed distributions, \ie, in a linear seed configuration (as \eg shown in Figure \ref{fig:tests2d}) limiting the diffusion distance to the closest $n$ other points is not sufficient, as the final cells do not limit each other but rather stretch out over the entire domain. In order to obtain region boundaries, additional postprocessing which includes subdivision and topological cleanups is required.
In contrast, our approach grows all cells concurrently and thus, stops the growth exactly and only when a cell's growth is blocked along its entire boundary.

\section{Layered fields}
\label{sec:layeredFields}

\label{sec:motivation}
As a motivating example, consider the problem of describing the state $\varphi$ of a melting ice cube.
The shape of the ice is continuously evolving over time. Modeling the interface across the liquid and solid states as a sharp boundary gives rise to several numerical challenges pertaining to discontinuities and to tracking the exact location of the boundary at each time step. Instead, when the boundary between the two states is defined as a narrow band where the function $\varphi$ describing the evolution gradually varies, these problems can be avoided and the location of the boundary is simply inferred from the state function~\cite{Fix:1981:PFM}.

In our context, the evolution of a growing region, or in other words, the position of its boundary, as a function of time, is implicitly given by the evolution of its state function $\varphi$ which is set to $1$ inside the region, $0$ outside and it is continuously varying as it traverses the narrow boundary band separating them as illustrated in Figure~\ref{fig:fieldprofile}. An important advantage of this formalism is that there is no need to track the boundary, \ie, to explicitly describe its location via mathematical equations, during evolution. In this way, numerical issues pertaining to sharp boundaries such derivative continuity are avoided in the first place.

\subsection{Derivation}

\begin{figure}
	\centering
	\includegraphics[width=\linewidth]{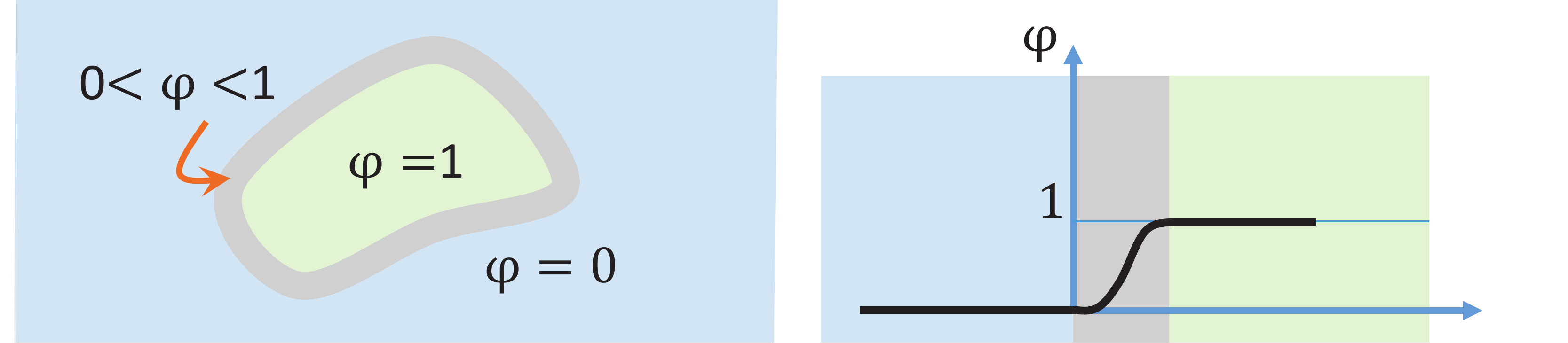}
	\caption{Definition of the state function over a region, and its profile as it crosses through the boundary (grey) of the region (green).}
	\label{fig:fieldprofile}
\end{figure}

\begin{figure}
	\centering
	\includegraphics[width=.78\linewidth]{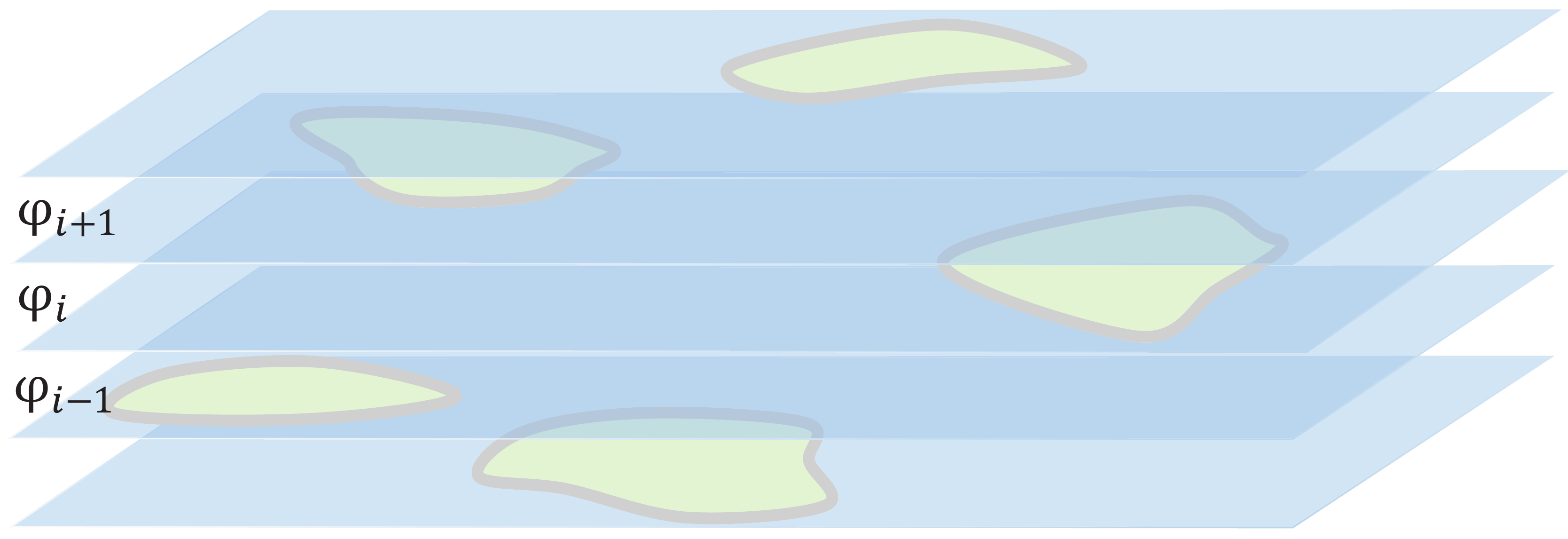}
	\caption{Illustration of the concept of layered fields. Each field evolves on a different layer, in this way, cell information is easily accessible by querying the corresponding layer.}
	\label{fig:layersMotivation}
\end{figure}

We regard the problem of tessellation formation as a growth process initialized at a set of given seeds (or initial regions) and driven by the evolution of the growing regions boundaries (interface). In view of Figure~\ref{fig:fieldprofile}, the interface can be described as a set of evolving narrow bands whose motion is consistent with the mechanism that governs the overall growth process. It is therefore possible to describe the entire tessellation by a single function, and to extract the interface after equilibrium as the values confined strictly between $0$ and $1$. Our exploration of this direction revealed that extracting the individual cells after convergence requires additional intricate geometric and topological operations. Instead, we describe the state of each cell by associating a function $\varphi_i$ with each seed $i$. In this way, we can capture the state of the entire tessellation while having easy access to the individual cells as illustrated in Figure~\ref{fig:layersMotivation}.

More formally, assume we have a set of $n$ evolving regions, initialized by seeds (sites), over a given surface. We associate a function $\varphi_i$ with each region. We define $\varphi_i$ as a function which takes value $1$ inside the cell and $0$ outside the cell. At the front, we have $0 < \varphi_i <1$, see Figure~\ref{fig:fieldprofile} for an illustration of the case of a planar surface.

In the following we establish some rules which govern the interaction between the fields $\varphi_i$ across the different layers. From the natural consideration that at equilibrium, any given point on the surface can only be inside a single region, or the narrow band separating different regions, we impose a partition of unity condition at all times
\begin{equation}\label{equ:phiSumto1}
\sum_{i=1}^n \varphi_i = 1.
\end{equation}
Clearly before equilibrium is reached, there will be surface areas that are neither inside a growing region or on narrow band, for this reason we introduce a base layer which is initialized to $1$ for all points. When seeds are initialized, the corresponding locations in the base layer are set to $0$.

\begin{figure*}
  \centering
  \includegraphics[width=.18\linewidth]{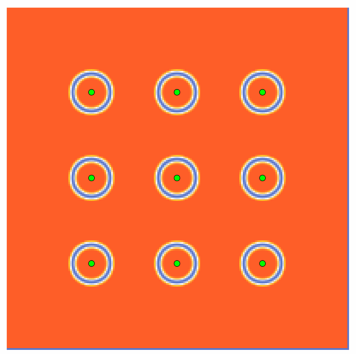}
  \includegraphics[width=.18\linewidth]{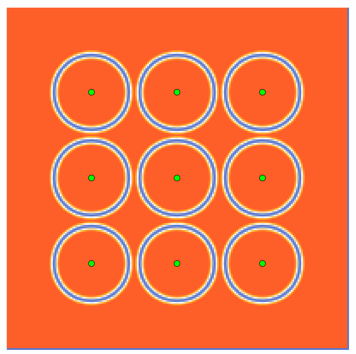}
  \includegraphics[width=.18\linewidth]{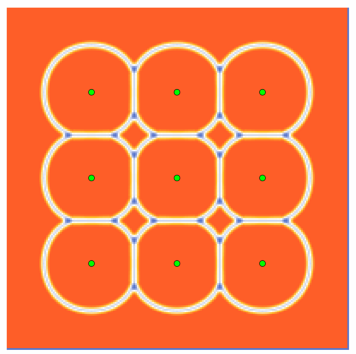}
  \includegraphics[width=.18\linewidth]{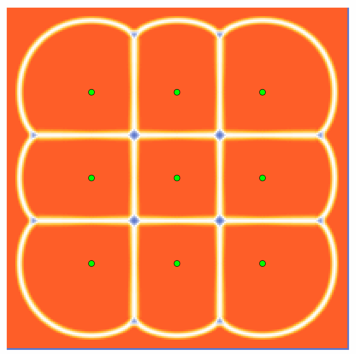}
  \includegraphics[width=.18\linewidth]{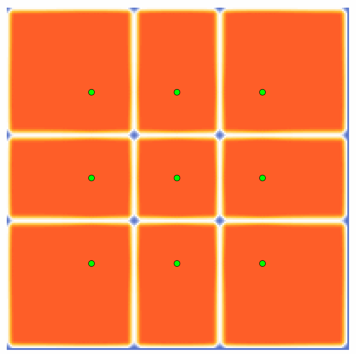}
  \includegraphics[width=.18\linewidth]{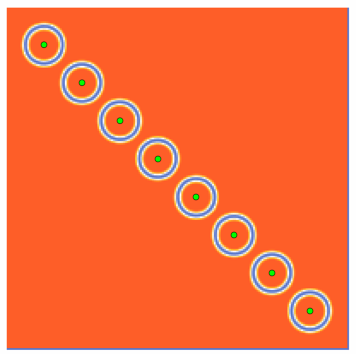}
  \includegraphics[width=.18\linewidth]{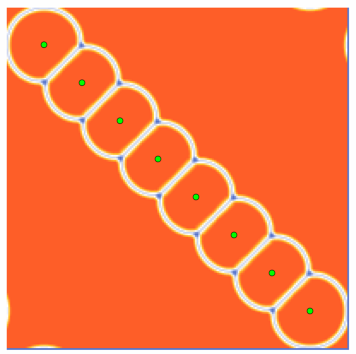}
  \includegraphics[width=.18\linewidth]{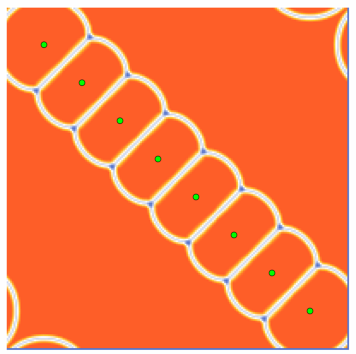}
  \includegraphics[width=.18\linewidth]{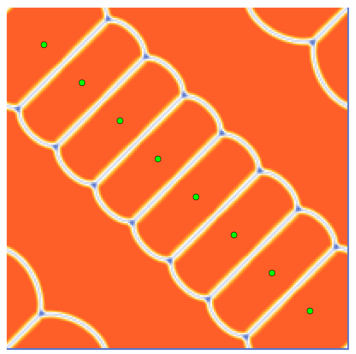}
  \includegraphics[width=.18\linewidth]{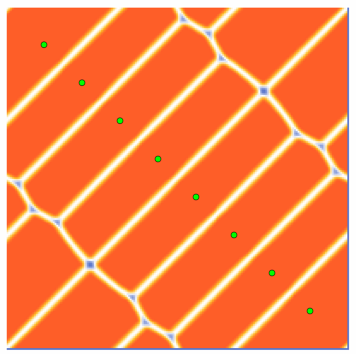}
  \caption{The progress of the propagation from seeds on two different planar configurations with periodic boundary conditions. Note that the effect of the $g$ as shown in Figure~\ref{fig:gplot} is visualized in blue.}
  \label{fig:tests2d}
\end{figure*}

In order to drive the evolution of our model, we need to define its energy.
Although the regions evolve on different layers, \cf Figure~\ref{fig:layersMotivation}, their interiors should not spatially overlap to reflect that cells are not allowed to merge into a single cell. This can be regarded as the fields associated with different cells are mutually exclusive. It ensues that we need to account for the products $\varphi_i \varphi_j$ for equilibrium.

Since the state functions of type $\varphi_i$ are constant within a region, they have null gradients and do not drive the evolution of the region. In contrast, the gradients on the narrow boundary bands do, \cf profile in Figure~\ref{fig:fieldprofile}. To reflect the condition that a growing region should not be allowed to change the shape of an already existing region, we can, similarly to above reasoning, capture the kinetics of the cell boundaries through the dot products of their gradients. In other words, we need to account for $\nabla\varphi_i \nabla \varphi_j$ for equilibrium.

It remains to describe the behavior of the system when two boundary bands meet.
Let the corresponding energy contribution be denoted as $g$ for now.
We will define it shortly.

From the discussion above, we define the global energy of the model on the surface $S$ as a combination of three terms,
\begin{equation} \label{equ:energy}
F  = \int_S f ds=\int_S \sum_{i=1}^n \sum_{j=i+1}^n (w_{ij} \varphi_i\varphi_j -\frac{a_{ij}}{2}\nabla\varphi_i \nabla \varphi_j + g) ds,
\end{equation}
where the penalty $w_{ij}$ and gradient energy $a_{ij}$ terms are fixed scalars. In all our experiments (we keep a ratio of 5, with wij=.2). We foresee them as parameters that could be used to adapt our approach to natural processes such biological cell growth.

By introducing the Lagrange multiplier $\lambda$ to account for the constraint of equation~\ref{equ:phiSumto1}, the $\varphi_i$s can be treated as independent variables, and the Lagrangian then reads
\begin{equation}\label{equ:Lagrange}
    \Gamma=\int_S f+\lambda (\sum_{i=1}^n \varphi_i -1) ds.
\end{equation}
The progression of $\varphi_i$ in the direction of the minimum of $\Gamma$ can then be expressed as follows
\begin{equation}\label{equ:timeEvolPhi}
    \mathring{\varphi_i}=-\frac{\partial\Gamma}{\partial \varphi_i}=-\frac{\partial F}{\partial \varphi_i}-\lambda.
\end{equation}
In the above notation, we consider only the direction and we drop the time scale (for now).
In order to neutralize the Lagrange multiplier, consider the following intermediate function
\begin{equation}\label{equ:diffPhi}
    \phi_{ij}=\varphi_i -\varphi_j; \quad \quad  (i<j).
\end{equation}
Clearly, $\phi_{ij}=-\phi_{ji}$. Substituting into equation~\ref{equ:phiSumto1}, we obtain
\begin{equation}\label{equ:phiPhi}
    \varphi_i=\frac{1}{n}(\sum_{j=1}^n \phi_{ij}+1).
\end{equation}
As the time evolution of $\phi_{ij}$ can be written in virtue of equation~\ref{equ:timeEvolPhi} as
\begin{equation}\label{equ:dotPhi}
    \mathring{\phi_{ij}}=\mathring{\varphi_i}-\mathring{\varphi_j}=-\frac{\partial F}{\partial \varphi_i}+\frac{\partial F}{\partial \varphi_j}.
\end{equation}
it follows that the time evolution of $\phi_{ij}$ is independent of the Lagrange multiplier $\lambda$.
From equation~\ref{equ:phiPhi}, we obtain
\begin{equation}\label{equ:dotphi2}
    \mathring{\varphi_i}=-\frac{1}{n} \sum_{j=1}^n (\frac{\partial F}{\partial \varphi_i}-\frac{\partial F}{\partial \varphi_j}).
\end{equation}
By considering a time scale (\ie, switching to standard time derivatives), equation~\ref{equ:dotphi2} can be written as
\begin{equation}\label{equ:phiTimeDerivative1}
    \dot{\varphi}=-\frac{1}{n} \sum_{j=1}^n \mu (\frac{\partial F}{\partial \varphi_i}-\frac{\partial F}{\partial \varphi_j}),
\end{equation}
by means of constant mobility term scalar $\mu$.

On the other hand, the functional derivative ${\partial F}/{\partial \varphi_i}$ reads
\begin{equation}\label{equ:functionalDerivative}
\begin{split}
{\partial F}/{\partial \varphi_i} &= \frac{\partial f}{\partial \varphi_i} - \nabla \frac{\partial f}{\partial \nabla\varphi_i}\\
                                  &= \sum_{\substack {j=1 \\j \neq i}}^n (w_{ij}\varphi_j+a_{ij}^2 \nabla^2 \varphi_j) +\frac{\partial g}{\partial \varphi_i}.
\end{split}
\end{equation}
Please note that in the above expression, the vector in the denominator (gradient in this case) is a standard shorthand notation for directional derivatives.
By means of equations~\ref{equ:phiTimeDerivative1} and \ref{equ:functionalDerivative}, we obtain
\begin{multline}\label{equ:phitimeDerivativeMain}
\dot{\varphi_i}=-\sum_{j=1}^n \frac{\mu}{n}\biggl(\sum_{k=1}^n\left[(w_{ik}-w_{jk})\varphi_k + \frac{1}{2}(a_{ik}^2-a_{jk}^2)\nabla^2 \varphi_k) \right] \\+(\frac{\partial g}{\partial \varphi_i}-\frac{\partial g}{\partial \varphi_j})\biggr).
\end{multline}
\begin{wrapfigure}{r}{0.45\linewidth}
\vspace{-10pt}
\hspace{-12pt}%
\includegraphics[width=1.1\linewidth]{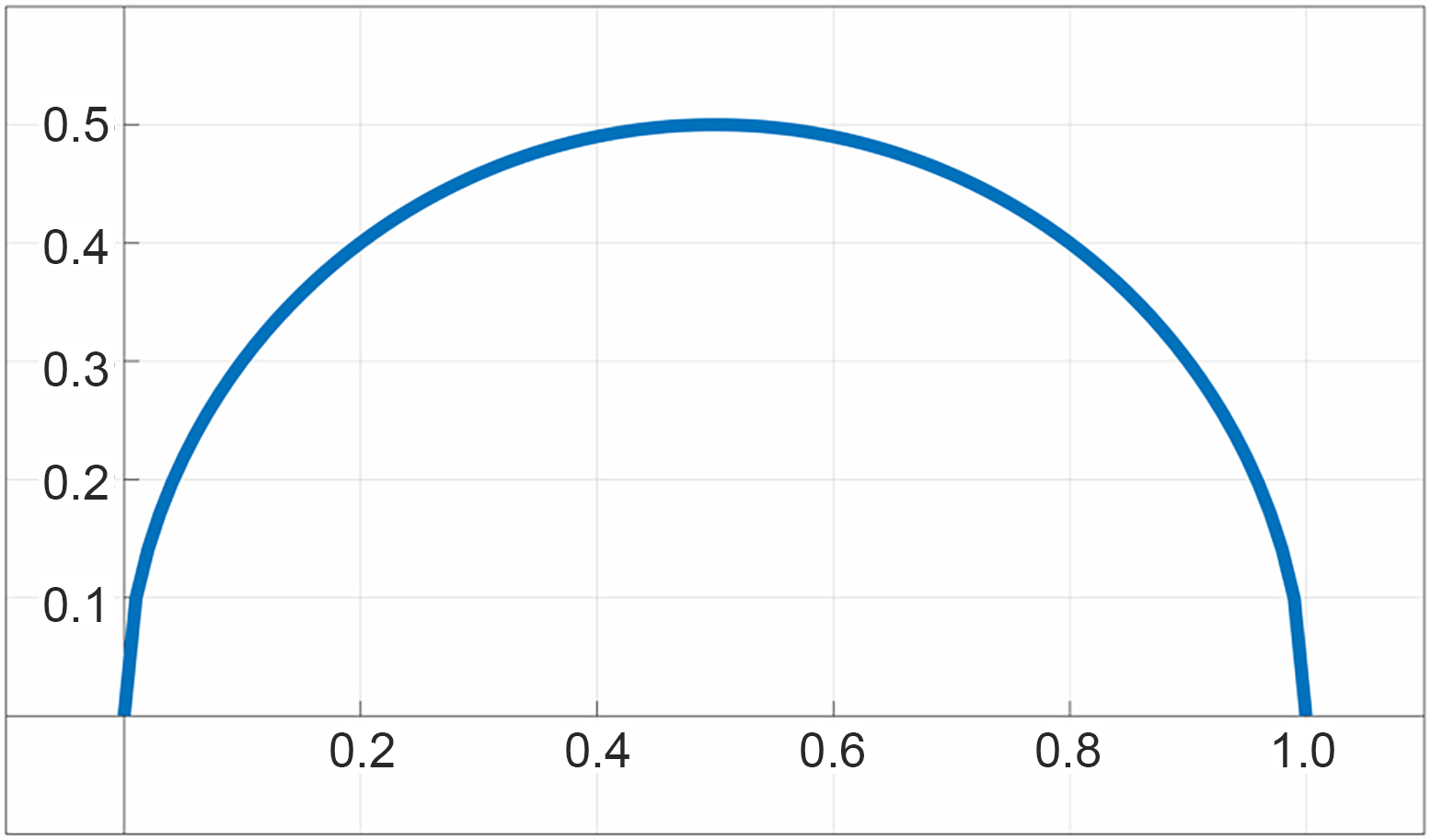}
\vspace{-15pt}
\hspace{-18pt}%
\caption{Plot of  $\sqrt{\varphi_i (1-\varphi_i)}$. }
\label{fig:gplot}
\vspace{-15pt}
\end{wrapfigure}
From the expression above, it turns out we do not need to define $g$ explicitly. Instead, we can directly define the expression of the difference term $(\frac{\partial g}{\partial \varphi_i}-\frac{\partial g}{\partial \varphi_j})$. To motivate our choice, we recall that $g$ encodes the desired behavior when two boundary bands meet each other. In the simple case where only two bands ${\varphi_i, \varphi_j}$ meet, we have by virtue  of equation \ref{equ:phiSumto1}, $\varphi_j=1-\varphi_i$. In practice , we seek a function that produces a symmetric behavior in  the interval $(0,1)$ and vanishes at $0$ and $1$. One such function is $\sqrt{\varphi_i (1-\varphi_i)}$ which has a suitable profile as illustrated in Figure \ref{fig:gplot}.

In this way, we define the remaining term
\begin{equation}\label{equ:dfe}
    \frac{\partial g}{\partial \varphi_i}-\frac{\partial g}{\partial \varphi_j} = -\sqrt{\varphi_i\varphi_j} e_{ij}
\end{equation}
where $e_{ij}$ is a constant scalar which describes the strength of interaction between the two boundary bands.

\begin{figure}
	\centering
	\includegraphics[width=.48\linewidth]{alignedrows3_layered_}
	\includegraphics[width=.47\linewidth]{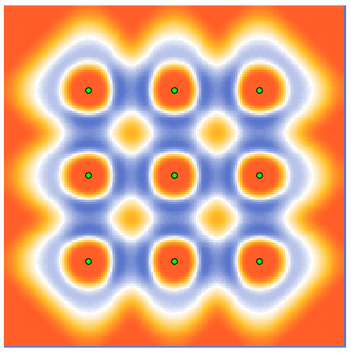}
	\caption{Effect of omitting the $g$ term in Equation \ref{equ:phitimeDerivativeMain}, would remove the separation between the fields.}
	\label{fig:tests2daltered}
\end{figure}

\begin{figure}
	\centering
	\includegraphics[width=.48\linewidth]{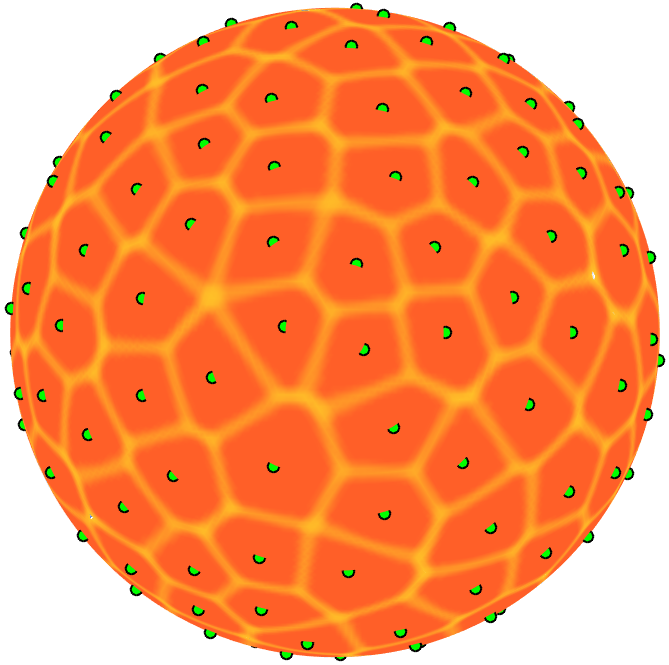}
	\includegraphics[width=.48\linewidth]{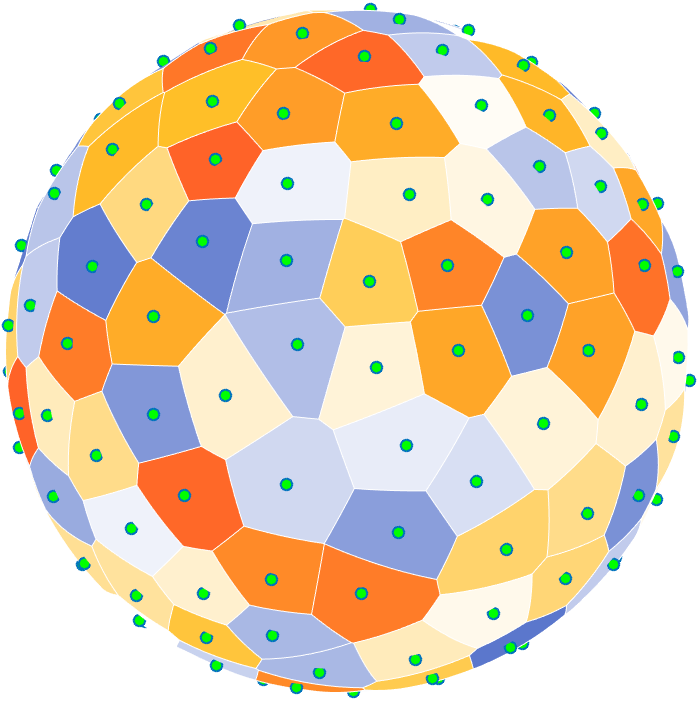}
	 \includegraphics[width=.49\linewidth]{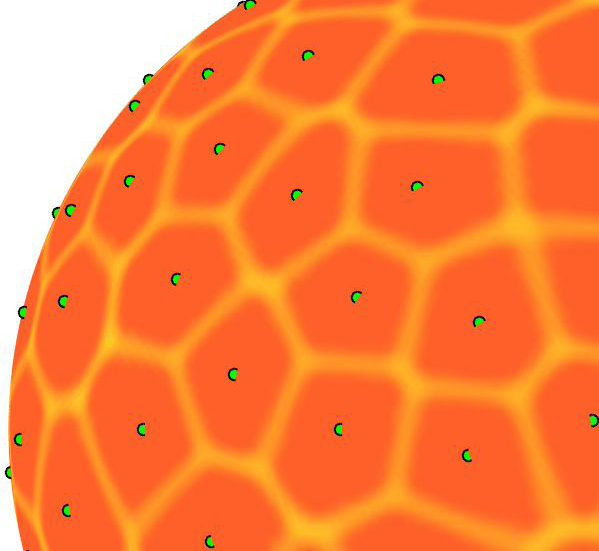}
	\includegraphics[width=.48\linewidth]{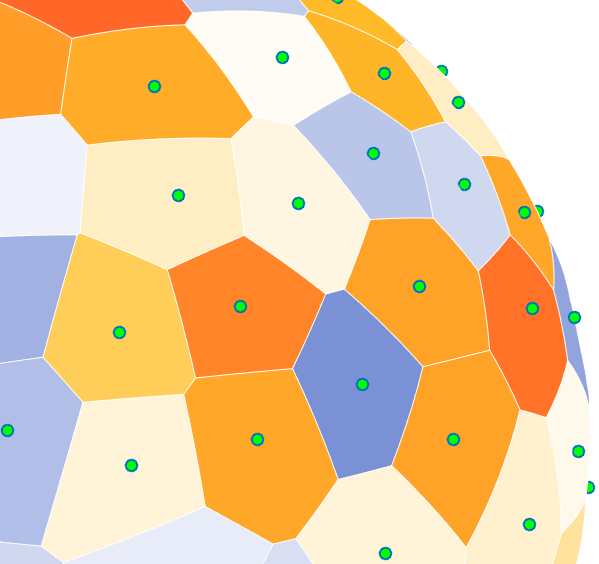}

	\caption{Side by side comparison of our approach (left) with the convex hull based Delaunay triangulation approach to Voronoi diagrams on the sphere~\cite{Renka:1997:ASD} (right). The original seeds are shown in green. A zoom-in reveals the quasi-similarity of the results.}
	\label{fig:testsSphere}
\end{figure}

This completes the definition of the time derivative of the governing equation of the system. The solution can then be carried by a simple explicit Euler stepping scheme
\begin{equation}\label{equ:Eulerstep}
    \varphi_i(t+\Delta t)= \varphi_i+ \dot{\varphi_i} \Delta t.
\end{equation}
To provide a first intuition of our approach, we present initial experiments for two different seed configuration in the planar setting in Figure~\ref{fig:tests2d}.
The experiments are in accordance with what would be obtained using Euclidian metric for partitioning and confirms the well behaved nature of our simple approach.
The effect of not accounting for gradient interactions at the narrow bands is shown in Figure~\ref{fig:tests2daltered}.
Extending our approach to the spatial setting, we focus on the particular case of the sphere where a Delaunay triangulation for a set of points laying on the sphere can be gained through computing their convex hull~\cite{Renka:1997:ASD}.  Starting from the same initial seed configuration shown as the green dots in Figure~\ref{fig:testsSphere}), the partitioning obtained with our approach is quasi-similar to the one from convex hull based approach and confirms the meaningfulness of our approach. For visualization, we use the squared sum of the field values at every vertex.

\subsection{Efficient problem representation}
\label{sec:algorithmicRepresentaiton}

In this section, we devise a concise and efficient representation of the problem geared towards simple implementation and portability. We represent the fields as a sparse matrix of size $(n_r \times n_v)$, where $n_r$ is the number of seeds or partitions and $n_v$ the number of vertices in the whole mesh. For each vertex $v_i$, we associate the corresponding values of $\varphi$s across the layers, \ie, $\varphi_{v_i}=({\varphi_1}_{v_i},{\varphi_2}_{v_i}, \dots,{\varphi_{n_r}}_{v_i})$. Since a vertex will either belong to a single region, will lie on the front across two different regions, or even at a junction (triple, quadruple, \dots), $\varphi_{v_i}$ is mostly populated by zeros. With this remark in mind, we represent the whole multi-layered field as the sparse matrix
\begin{equation}\label{equ:globalphi}
    \mathbf{\Phi}=(\varphi_{v_1}, \varphi_{v_2}, \dots,\varphi_{v_{n_v}} ).
\end{equation}

From a data querying perspective, this formalism offers an elegant way of extracting topological information. Vertices within a cell are the ones that flag non-zero values along the row in $\mathbf{\Phi}$ that corresponds to the cell.

From an implementation perspective, this formalism simplifies the algorithmic pipeline to a great extent. Since the geometry of the mesh is static we pre-compute the Laplacian $\mathbf L$ of the mesh only once. At each iteration, the evaluation of the second term of equation~\ref{equ:phitimeDerivativeMain} on the whole surface and over all layers amounts to a single sparse matrix-matrix multiplication, \ie, $\mathbf{\Phi L^{\top}}$.

The algorithmic outline of our approach is shown in Algorithm~\ref{alg:layeredfields} and can be broken down into four major steps:
\begin{enumerate}
	\item initialization based on seeds or regions (ln 1)
	\item Laplacian computation  and evaluation of the layered field Laplacian (ln 2-4)
	\item update of the layered field according to equation~\ref{equ:phitimeDerivativeMain} (ln 7-14)
	\item normalization (ln 20)
\end{enumerate}

\begin{algorithm}
	\DontPrintSemicolon
	\SetKw{To}{to}
    Initilize $\mathbf{\Phi}$ with seeds and/or initial regions\;
    Precompute Laplacian $L$\;
	\For{$t$ $\leftarrow$ $t_0$ \To $t_{end}$}{
		$L_{t}=\mathbf{\Phi} L^{\top} $\;
		\For{$i\leftarrow 1$ \To  $n_v$}{
            $\text{irows} =(\mathbf{\Phi}_{\star,i}>0) \lor ( (\mathbf{\Phi}_{\star,i}=0) \land ({L_t}_{\star,i}>0))$\;
            $n_i=\mathopen|\text{irows}\mathclose|$\;
            \For{$j\leftarrow 1$ \To  $n_i$}{
                $d \leftarrow 0$\;
                \For{$k\leftarrow 1$ \To  $n_i$}{
                    $\text{sum} \leftarrow 0$\;
                    \For{$l\leftarrow 1$ \To  $n_i$}{
                    $\text{sum}+ =  \frac{1}{2} (a_{j,l}-a_{k,l}) L_{t}(l,i)+(w_{j,l}-w_{k,l}) \mathbf{\Phi}(l,i)$\;

                    }
                    $d+= - \frac{\mu}{n_i}\biggl(sum- e_{j,k} \sqrt{\mathbf{\Phi}(j,i) \mathbf{\Phi}(k,i)}\biggr)$\;

                }
                $\mathbf{\Phi}(j,i)+=d~ \Delta t$\;

                \If{($\mathbf{\Phi}(j,i)>1$)}{
                 $\mathbf{\Phi}(j,i)=1$\;
                 }
                \If{($\mathbf{\Phi}(j,i)\leq 0$)}{
                 $\mathbf{\Phi}(j,i)=0$\;
                 }
            }
		}
		Normalize each column of $\mathbf{\Phi}$ by the sum of its nonzeros\;
	}
	\caption{Multi-layred field algorithm}
	\label{alg:layeredfields}
\end{algorithm}

\paragraph{Initialization}
Our approach can be initialized by setting the values of the field $\mathbf\Phi$ at the corresponding vertices to $1$.  Ideally, a vertex and its immediate neighbors can be set to $1$ or to a function defined on its neighboring triangles. There is no need at this stage to accommodate the boundary of the seeded regions to account for the narrow band of the interface.
Please note, that energy defined earlier accounts mainly for interaction between cells. In order to allow cells to evolve freely beforehand, we introduce an additional base layer as mentioned earlier in Section~\ref{sec:layeredFields}. The corresponding $\varphi_b$ evaluates to $1$ at all non-seeded vertices.

\paragraph{Laplacian computation}
Since the mesh is static, we precompute the Laplacian of the mesh and store the resulting sparse matrix. Since the Laplacian is not updated throughout the rest of our algorithm, the cost of this operation does not affect the overall iterative process. Nonetheless, within the time loop the Laplacian of the phase field needs to be updated (ln.4.), the cost of this operations amounts to a sparse matrix-matrix multiplication.

\paragraph{Update}
The main operation within the time loop is to update the layered field $\mathbf{\Phi}$. This spans lines $5-20$. The first step is to detect the growing regions, that is regions where the field is nonzero as well as possible areas that can get affected by the propagation, \ie, regions where the field Laplacian is positive. Both operations are performed in ln.~$6$. The output of this operation is \texttt{irows} which indicates the active rows for the field evaluated at vertex $i$. For the columns $\varphi_i$, we iterate over the rows of interest, \ie, \texttt{irows} and perform a summation over the entries of the corresponding field Laplacian multiplied by a factor which encodes the nature and the strength of interactions across the involved layers. Clearly, this operation is akin to sparse matrix-matrix addition with additional scaling for the involved quantities. Once this columns addition is completed, a cleaning operation needs to be performed to keep the field within the defined range, namely, above zero and below one.

\paragraph{Normalization}
Once the field update is computed, a normalization is performed to enforce that the sum of field values across layers at every vertex equals $1$.

From the discussion above, the cost of one time step is dominated by two operations and can be regarded as the cost of sparse matrix-matrix multiplication and a sparse matrix-matrix addition. The latter is in fact relatively negligible. Given the pure algebraic nature of our approach it can be easily ported to modern parallel computing platforms. An outline of our  implementation on the graphics computing unit (GPU) is given in the following section.

\section{GPU Implementation}
\label{sec:GPU}
The major challenge of a GPU implementation is finding a good way to parallelize the entire algorithm.
The first step in each iterations is a simple sparse matrix-matrix multiplication, for which a variety of GPU libraries exist.
Mostly iterating over the matrix columns, we use a compressed sparse column (CSC) representation, which describes a matrix using a linear array of values and associated row indices, as well as a column pointer array that points to the beginning of every column.
The second step, the update, consists of four nested loops (ln 5, 8, 10, 12). The two outer loops describe updates of individual entries of a sparse matrix (ln 15).
As these operations are independent of one another, these loops form a major potential for parallelization.
From a high-level perspective this corresponds to a sparse matrix-matrix addition.
However, this update is complicated by the inner loops (ln 10 and 12), which each require serial iterations over entire matrix columns.
Also, the parallelization of the outer two loops is non trivial, as the loops run over the entries of interest, which vary between vertices and also change from iteration to iteration.
The final step of each iteration, the normalization, can obviously be parallelized over each column.

\paragraph{Laplacian}
Although sparse matrix-matrix multiplication is algorithmically the most complicated step on a parallel device like the GPU, we can make use of readily available libraries like cuSparse~\cite{Nvidia:cusparse}.
Internally, cuSparse computes an explicit transpose for $L^T$, which can take up nearly half the overall time spent on the multiplication.
As we consider the Laplacian of the mesh constant throughout all iterations, we can precompute the transpose once, and reuse it throughout all iterations.

The Laplacian of the phase field $L_t$ is computed in every iteration. Allocating memory for it in every iteration would lead to a significant overhead.
Thus, we allocate memory before starting the iterations and reuse it in every step.
However, $L_t$ can obviously grow from iteration to iteration.
Thus, in case the preallocated storage is not sufficient for $\mathbf L_t$, we have to allocate it anew.
To provide enough storage for multiple iterations, we enlarge it by $20\%$ at once.

\paragraph{Update}
In order to parallelize the for loops involved in the update routine, we first determine the values of $i$ and $j$ throughout the entire update.
To this end, we need to identify all rows of interest (\texttt{irows}) for all vertices.
We do this by evaluating ln 6 of the algorithm in parallel for all vertices (columns of the involved matrices).
The result of this operation corresponds to a vector \texttt{irows} for every vertex, which as a whole corresponds to a sparse matrix skeleton---describing where rows of interest are, without assigning any value to them.
We call this matrix skeleton the \emph{interest skeleton}.
The construction of the skeleton involves a prefix sum~\cite{Sengupta:2007:SPG} over the number of entries in every column.
Memory needs to be allocated for the skeleton in every iteration. We again, follow the same strategy as before, allocating memory once and enlarging it by $20\%$ in case it is not sufficient.

After the interest skeleton has been constructed, we transfer $\mathbf{\Phi}$ and $L_t$ into a representation that matches the skeleton, adding explicit zero entries for entries not present in the original matrices.
We call these representation $\hat{\mathbf{\Phi}}$ and $\hat{L}_t$.
Although mathematically there is no difference between $\mathbf{\Phi}$ and $\hat{\mathbf{\Phi}}$, $L_t$ and $\hat{L}_t$, having all involved matrices within the same skeleton (and thus the same non-zero pattern), allows for more efficient parallelization of the following steps.

Our next concern is the innermost summation of the algorithm (ln 12-13). Analyzing its structure, it can be seen that it describes an interaction between the different fields at a given vertex.
When evaluating the influence of field $k$ on $l$ and field $l$ on $k$ equal terms are considered. A straight forward implementation would compute these respective continuations multiple times.
Taking a slightly different approach, we can avoid these duplicated computations.
The loop in line 12 with the partial sum contributions can be reinterpreted as the differences of selected results from the matrix-matrix multiplications
\[
A_L = a\cdot L_t \text{\hspace{18pt} and \hspace{18pt}} W_\mathbf{\Phi} = w\cdot \mathbf{\Phi}.
\]

Using this result, the summation (loop in ln 12) can be rewritten:
\begin{equation}
	sum = \frac{1}{2} (A_L(j,i) - A_L(k,i)) + (W_\mathbf{\Phi}(j,i)  - W_\mathbf{\Phi}(k,i)),
	\label{equ:alt_sum}
\end{equation}
whereas only selected results from the respective matrices are used.
As these used values correspond exactly to the entries present in the interest skeleton, we can compute them efficiently in parallel.
Starting one thread for every entry in the skeleton on the GPU, we assign each thread to the row $i$  querying its own row coordinate from the skeleton.
We then compute a dot product between a dense vector (row $i$ in $a$/$w$) and a sparse vector (the entire column associated with the current entry, representing ($l$)) to compute the resulting entry of $A_L$/$W_\mathbf{\Phi}$.

With $A_L$ and $W_\mathbf{\Phi}$ at hand, we compute the update for $\mathbf{\Phi}$ (ln 14) for all entries of $\hat{\mathbf{\Phi}}$ in parallel.
Again starting one thread of each entry in the interest skeleton, each thread runs through the loop over $k$ (ln 10).
We evaluate the sum according to Equation~\ref{equ:alt_sum} by fetching four values from $A_L$ and $W_\mathbf{\Phi}$ and compute the update to $d$ (ln 14) by fetching the respective values from $t$, $e$ and $\hat{\mathbf{\Phi}}$.
Note that  due to using $\hat{\mathbf{\Phi}}$ here, the lookup for $\mathbf{\Phi}(j,i)$ and $\mathbf{\Phi}(k,i)$ is trivial although the matrices are sparse, because all involved matrices follow the interest skeleton.
Furthermore, the individual entries $\mathbf{\Phi}(k,i)$ correspond to a single column in the interest skeleton and thus are right next to each other in memory.

The update for $\mathbf{\Phi}$ and clamping of the value (line $15-19$) can be completed alongside the computation of the update itself.
The output of this entire step is an updated copy of $\hat{\mathbf{\Phi}}$.

\paragraph{Normalization}
The normalization is carried out via two simple steps.
At first, we compute a sum over each column in parallel. In a second step, we launch one thread for each individual entry of $\hat{\mathbf{\Phi}}$ and update each value.
Working with $\hat{\mathbf{\Phi}}$ instead of $\mathbf{\Phi}$ may introduce additional zero entries in the result (note that ln 18 and 19 may also introduce explicit zero entries).
It is desirable to remove those entries to avoid unnecessary increases in the representation.
However, we do not have to perform this compaction step explicitly, as it is implicitly carried out by computing the interest skeleton in the next iteration.

After computing the final iteration, we perform an explicit compaction to avoid zero entries in the output. We combine this compaction with the normalization step, counting the number of non zero entries in every column alongside its sum.
We then again run a prefix sum to generate the compacted column pointer array as required by the sparse matrix representation and fill the compacted matrix while multiplying with the normalization factors.

\paragraph{Typical relative performance}
To provide an intuition about the costs of the described steps, we provide detailed timings for one iteration of the algorithm on the Bimba model depicted in Figure~\ref{fig:bimba} with $3.7M$ triangles and $2500$ seeds in Table~\ref{tab:perfbreakdown}. Clearly, the performance is dominated by the matrix-matrix multiplication. After computing the interest skeleton, the remaining steps can be parallelized very efficiently, leading to a negligible cost for the update itself.
\begin{table}[h]
\centering
\begin{tabular}{rcr}
	Matrix-Matrix Multiplication & 20.241ms & 84.3\%\\
	Interest Skeleton Creation & 1.321ms & 5.5\%\\
	Computing $\hat{L}_t$ $\hat{\mathbf{\Phi}}$ & 0.720ms & 3.0\%\\
	Updating  $\hat{\mathbf{\Phi}}$ & 1.177ms & 4.9\%\\
	Normalization & 0.552ms &  2.3\%\\ \midrule
	Sum & 24.011ms & 100\%
\end{tabular}
\caption{Performance breakdown for a single iteration of our algorithm on the Bimba model (\protect Figure~\ref{fig:bimba}) with $3.7M$ triangles and $5000$ seeds for an NVIDIA GTX Titan Xp GPU. Due to our optimizations on all other steps, the matrix-matrix multiplication clearly dominates performance.}
\label{tab:perfbreakdown}
\end{table}

\section{Lloyd's algorithm and mesh extraction}
\label{sec:approximation}
Adapting our approach to perform Lloyd-like iterations is straightforward. Starting from a random distribution of sites on the surfaces, we perform the first partitioning as outlined earlier. The vertices of each individual cell are simply the ones that flag nonzero entries in the row of the sparse matrix $\Phi$, corresponding to the cell layer. For performing center of mass update, one can resort to existing methods discussed in Section~\ref{related}. Interestingly though, we observed that even a very simple approximation of the centroid location works well with our approach. We compute the center of mass as the average of the barycenters of all triangles of the cell (have a vertex in the cell) weighted by their respective triangle area. Clearly, this center of mass does not necessarily lay on the surface. We back-project it on the cell along the vector obtained as the average of the triangle normals weighted by their respective area. Cells shapes can be complicated, and at times, the intersection can be more than a single point. In this case, we simply choose the closest to the center of mass. If there is a ray miss, we do not update the seed position at that particular iteration. In this way, cell-center computation does not require explicit distance computation and the overall procedure adheres to the simplicity sought in this work.

\begin{figure}[h]
	\centering
	\includegraphics[width=.98\linewidth]{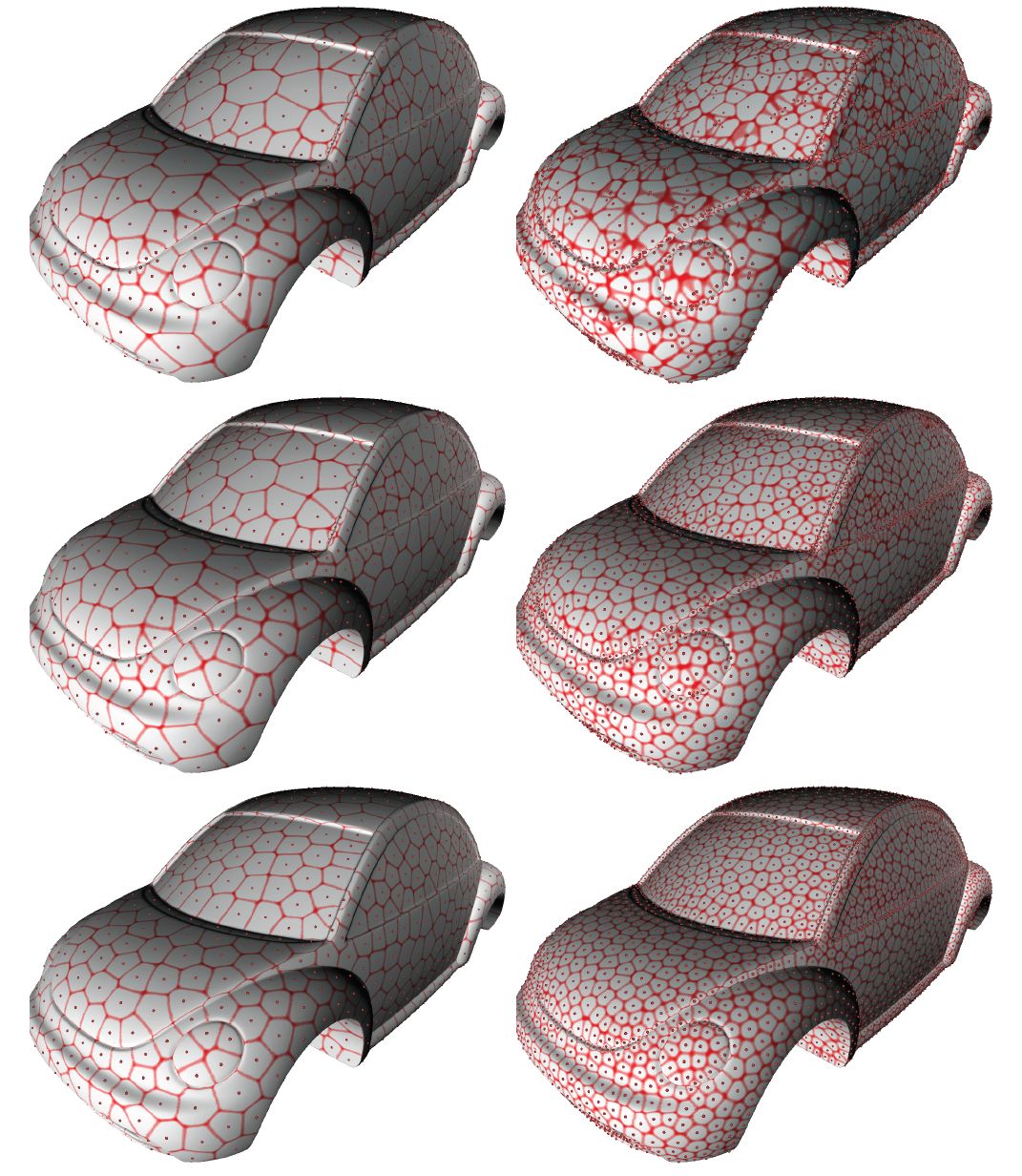}
	
	\caption{Lloyd-like iterations on the Beetle mesh (2M$\Delta$)  for $1000$ (left) and $5000$ (right) seed locations. The initial random distribution (first row) features extremely uneven clustering of seeds (right). The distribution becomes significantly better after a first approximate center of mass update (second row) and after 15 Lloyd-like iterations (last row).}
	\label{fig:beetle}
\end{figure}

\begin{figure}
	\centering
	\begin{subfigure}{\linewidth}
		\centering
		\includegraphics[width=.98\linewidth]{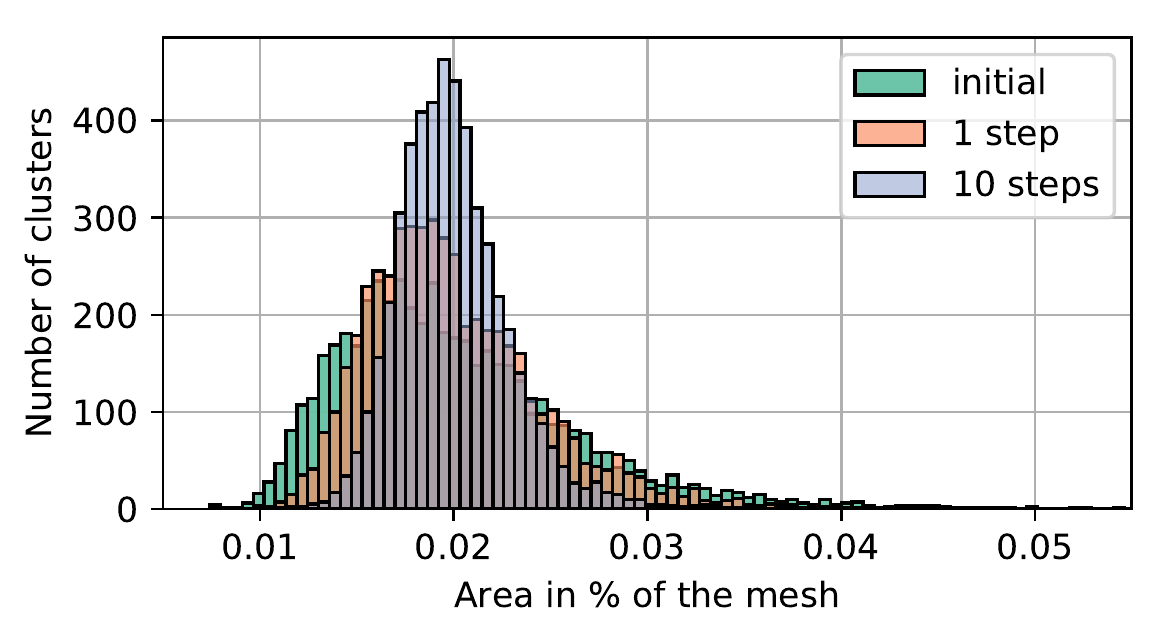}\\
		\vspace{-8pt}
		\caption{Bimba $5000$}
	\end{subfigure}
	\begin{subfigure}{\linewidth}
		\centering
		\includegraphics[width=.98\linewidth]{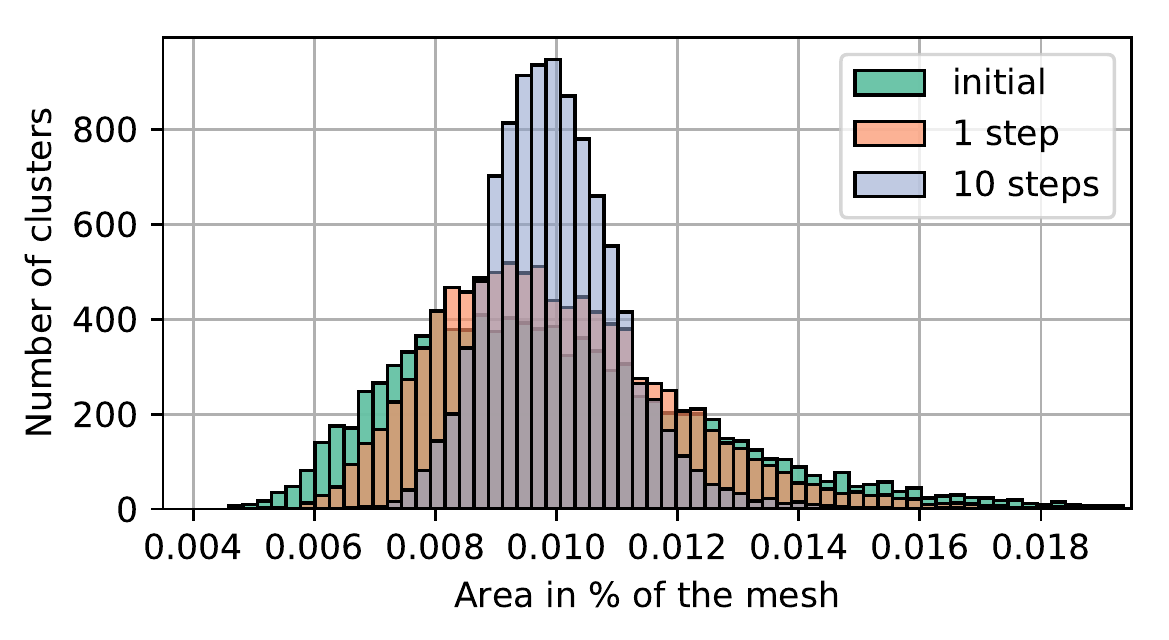}\\
		\vspace{-8pt}
		\caption{Flightsuit $10\,000$}
	\end{subfigure}\\
	\vspace{-8pt}
	\caption{Initial cell area histograms and the effect of $1$ and $10$ Lloyd iterations for the models shown in Figure~\ref{fig:bimba} and Figure~\ref{fig:flightsuit}.}
	\label{fig:areadistrib}
\end{figure}

While more elaborate schemes for center of mass estimation could be used, \eg, ~\cite{Peyre:2006}, this simple procedure works well in all our experiments, as shown in Figure~\ref{fig:beetle}, and serves the purpose of illustrating the feasibility of LLoyd's iteration with our layered-fields representation. The evolution of cell area distribution summarized in Figure~\ref{fig:areadistrib} confirms the well behaved nature of our approach.

\begin{figure}
 	\centering
 	\begin{subfigure}[b]{0.9\linewidth}
 	\centering
    \includegraphics[width=.8\linewidth] {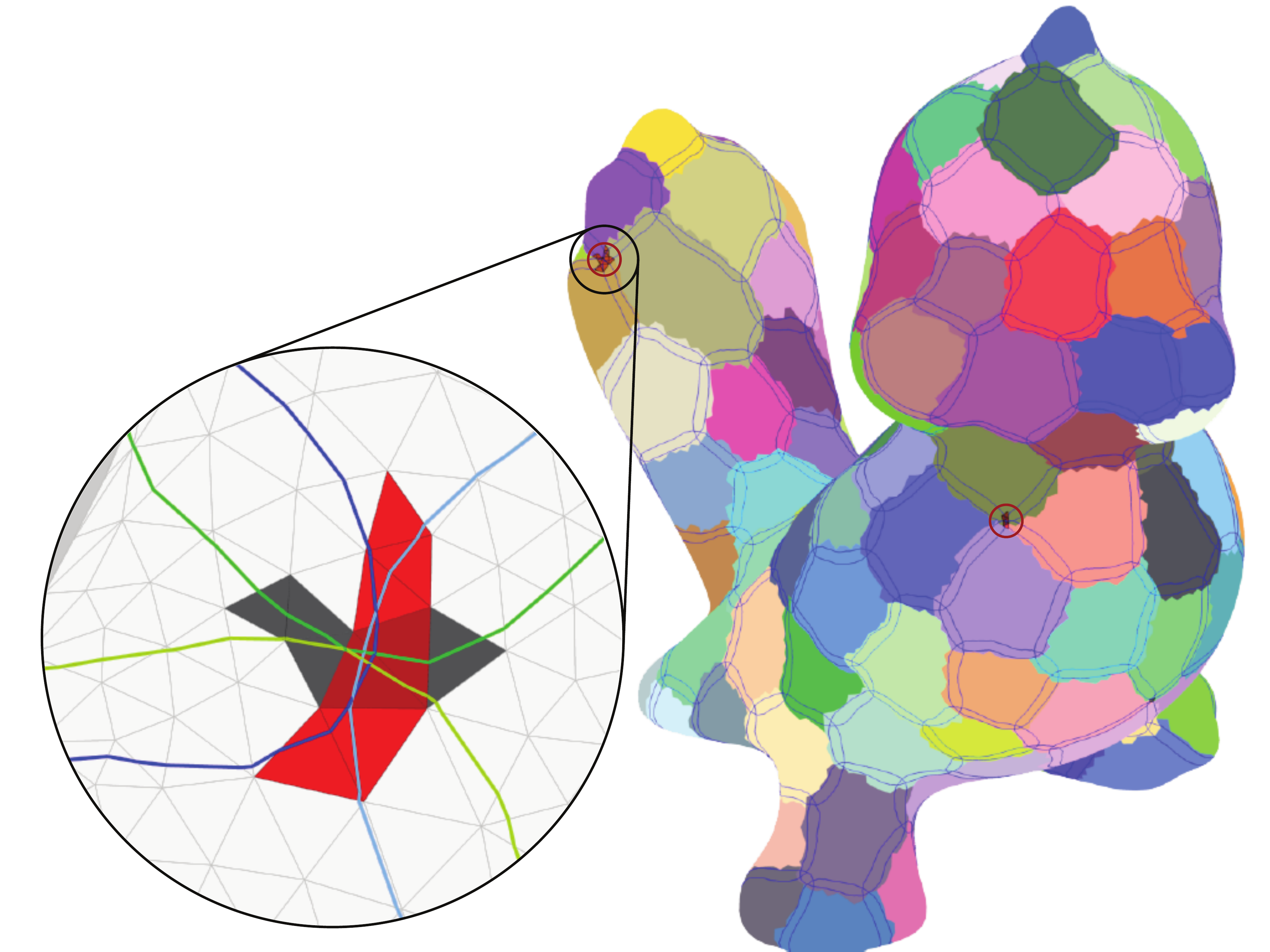}
    \caption{}
   	\label{fig:degenracies_cells_isos}
    \end{subfigure}
    \vspace{-20pt}

    \begin{subfigure}[b]{0.48\linewidth}
    \centering
	\includegraphics[width=.6\linewidth] {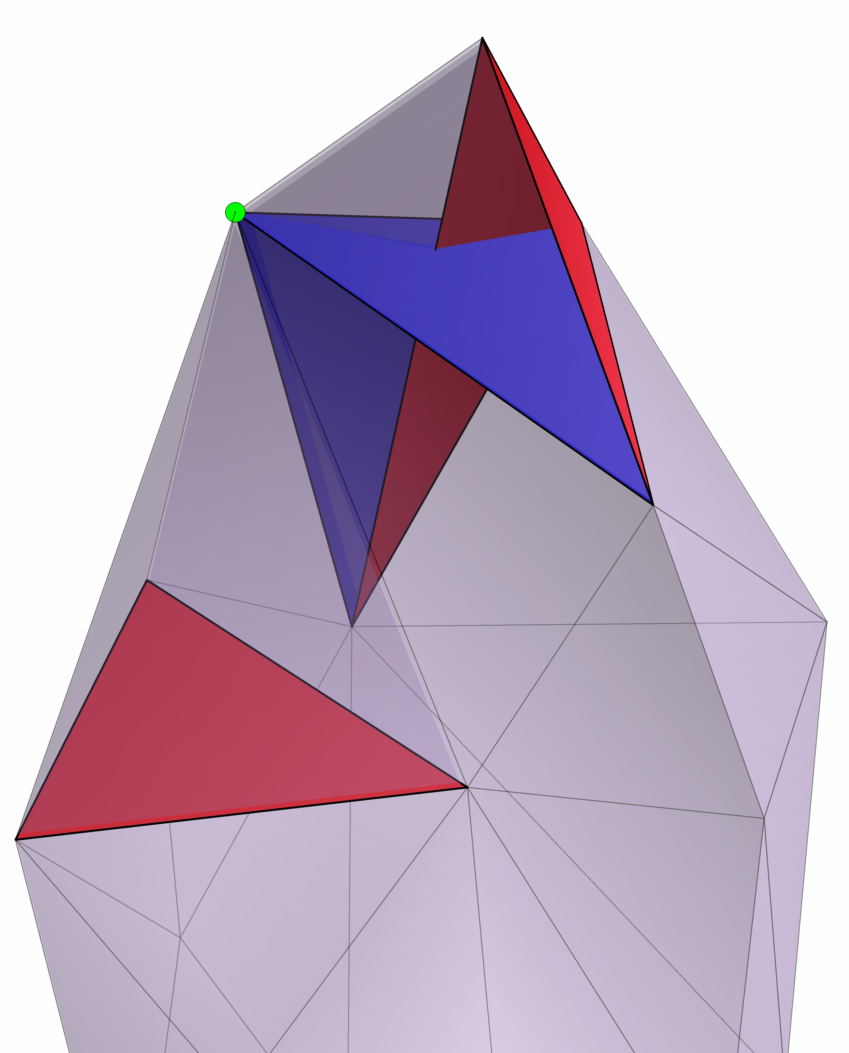}
	\caption{}
	\label{fig:degenracies_cells_folding}
	\end{subfigure}
    \begin{subfigure}[b]{0.48\linewidth}
   	\centering
    \includegraphics[width=.8\linewidth] {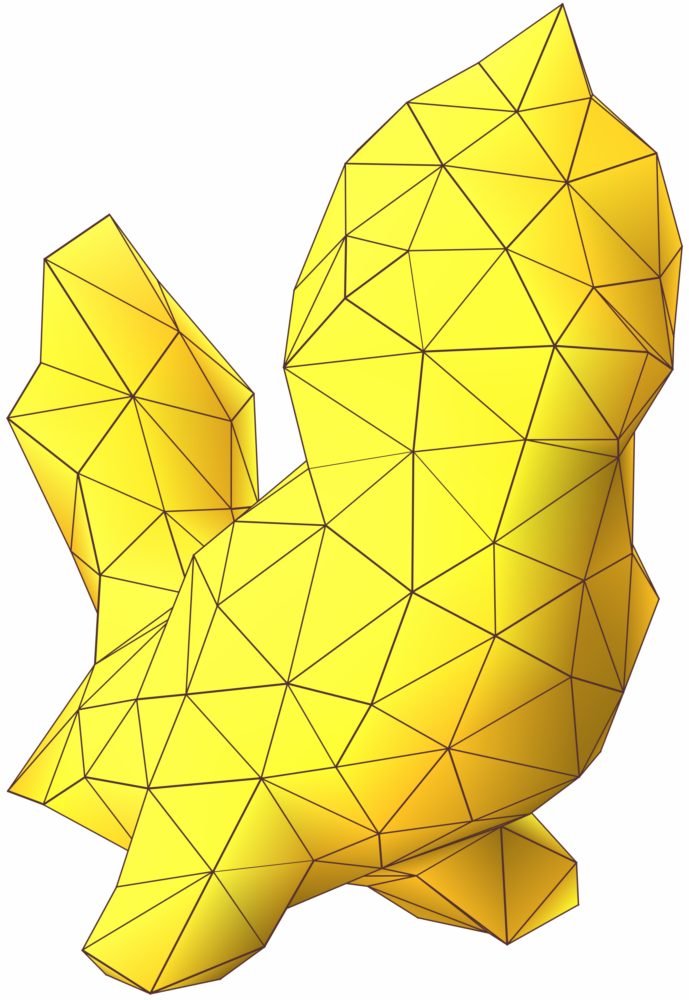}
    \caption{}
    \end{subfigure}
  \vspace{-10pt}
  \caption{(a) Segmented cells and their lax isolines with a threshold of $.25$ for the Tweety model. Some cell intersection only happen on triangles but not on vertices (red), with the zoom-in detailing one such intersection. (b) Even after exact isoline intersection, non-manifold meshes would be generated if more than three isolines intersect. (c) We prohibit this situation from happening by restricting the number of shared neighbors of pairs of potentially intersecting cells.}
  \label{fig:degenracies_cells}
\end{figure}

\subsection{Cell neighborhoods and dual mesh}
There are many ways for determining cell neighborhoods and extracting the dual mesh.
For instance, tracing along a given levelset of the function defined as the squared sum of $\Phi$ across the layers would be an option. In fact, the resulting isolines inside the narrow band can be used in a similar fashion to field-based remeshing methods, \eg,~\cite{Schall08}.
However, we found that our layered field representation offers a more malleable and computationally efficient alternative.

We define isolines per individual cells. In view of the definition in equation~\ref{equ:dfe}, a threshold value $.25$ or slightly lower for the field $\Phi$ would amply guarantee intersection of neighboring cell isolines and subsequently that their underlying triangulations overlap. An example of such isolines is shown in Figure~\ref{fig:degenracies_cells_isos} as blue closed curves (a much lower threshold would simply yield too much overlap). However, instead of performing an extensive isoline computation and intersection search over the whole mesh, we capitalize on the overlap itself to define the cell-cell adjacency relations.

In our formulation, the layered field is represented as a sparse matrix where each column represents a vertex and each row represents a cell. A preliminary estimation of the cell-cell adjacency matrix can be obtained by assuming that two cells are adjacent if they share a vertex. Such matrix can be obtained as $A_v= \bar{\Phi} \bar{\Phi}^\top$, with the diagonal set to zero. $\bar{\Phi}$ is the sparse matrix where values of $\Phi$ that pass the threshold are set to one and to zero otherwise. Unfortunately $A_v$ cannot capture all cases induced by the intersections of cell-cell isolines as two overlapping regions can intersect without sharing a vertex, as shown in the zoom in Figure~\ref{fig:degenracies_cells_isos}. Those cases need to be captured at the triangle level. To do so, we can obtain a different adjacency which encodes that cells are adjacent if their respective isolines pass on the same triangle. It can be obtained through sparse matrix algebra as well using a logical variant of $A_t= (\bar{\Phi} \bar{M}) (\bar{\Phi} \bar{M})^\top$, where $\bar{M}$ is the $n_v \times n_f$ face-vertex incidence matrix with $n_v$ and $n_f$ being the number of vertices and faces respectively. This is the sparse binary form of the recently introduced mesh matrix formulation~\cite{meshmatrix17}. Again for $A_t$ we set the diagonal to zero.

Clearly the nonzero entries of $A_v$ are a subset of those of $A_t$. Furthermore, entries which are unique to $A_t$ encode cases where two or more isolines travel the same triangle.
However, they do not reveal whether they effectively intersect. We can single out these cells using the nonzero entries of ($A_t-A_v$) and obtain the candidate triangles on which the isolines will be effectively computed and the intersection checks performed.
These triangles are obtained based on overlapping entries of the respective rows of the involved cells in the matrix ($\bar{\Phi} \bar{M}$). In practice, these constitute only a very small number of triangles. For instance, there are only two such cases (red circles) on the mesh in Figure~\ref{fig:degenracies_cells_isos}.
The zoom-in shows two sets of triangles (black and red) on and their corresponding  isolines (two shades of green and blues resp.). Once this check is performed, we can confirm whether the involved isolines intersect at those triangles by simple intersection checks. In this way, we can update the corresponding entries of $A_t$ according to the intersection scores. If there is no effective intersection the entry is set to zero. This adjacency matrix is not guaranteed to yield a manifold mesh as there could be more than three isolines that intersect at certain locations, as shown in example Figure~\ref{fig:degenracies_cells_folding}. As the treatment of such degeneracies is generally difficult in post processing, we overcome this issue by automatically enforcing the requirement on those candidates in the adjacency matrix itself. For the candidates in ($A_t-A_v$)---when an intersection is detected---we check if the two involved cells have two or more already confirmed shared neighbors. If so, we drop the additional intersection in $A_t$. Throughout this simple approach, we can ovoid complicated topological cleaning upfront.

\subsection{Mesh construction}
After the manifoldness check above, we can proceed to construct the dual mesh from the adjacency matrix $A_t$. To the best of our knowledge, existing graph based approaches require a planar embedding and then perform ring construction based on angle considerations. Our approach is simpler and is based on matrix algebra as described below. Staring from the curated version $A_t$, for a given vertex $i$, the corresponding vertex ring (neighbors) are the nonzero elements of row $i$ of $A_t$. Popping one elements from the ring, say vertex $j$, we simply intersect the neighbors of $j$ with the ring elements. This can be also algebraically encoded as the multiplication of the rows of $A_t$ corresponding to the ring elements with the vector $x_i$ which is set to zero everywhere and one at $i$. This step yields a vector with a maximum of two nonzeros entries, each of which forms a triangle with $i$ and $j$. We repeat this process till the ring is empty. Once all vertices have been processed an initial triangulation is obtained.

A typical case of spurious triangles may occur when a cell has only three neighbors which are also neighbors to each other (around a bump or a spike). These case can be easily discarded by finding rows of $A_d$ which contain only three entries and checking if these three entries form a triangle. This last step can be conveniently done by multiplying those rows by the binary mesh matrix of the newly constructed dual mesh. Rows in the output vector which flag a value of $3$ refer to the ids of spurious triangles which need to be removed from the triangulation. The last step in the dual mesh construction amounts to consistently orienting the triangles. This can be done in the customary way starting from an initial triangle and orienting its edge-neighbors consistently and repeating the process till all triangle have been processed.

\begin{figure}
		\includegraphics[width=.98\linewidth] {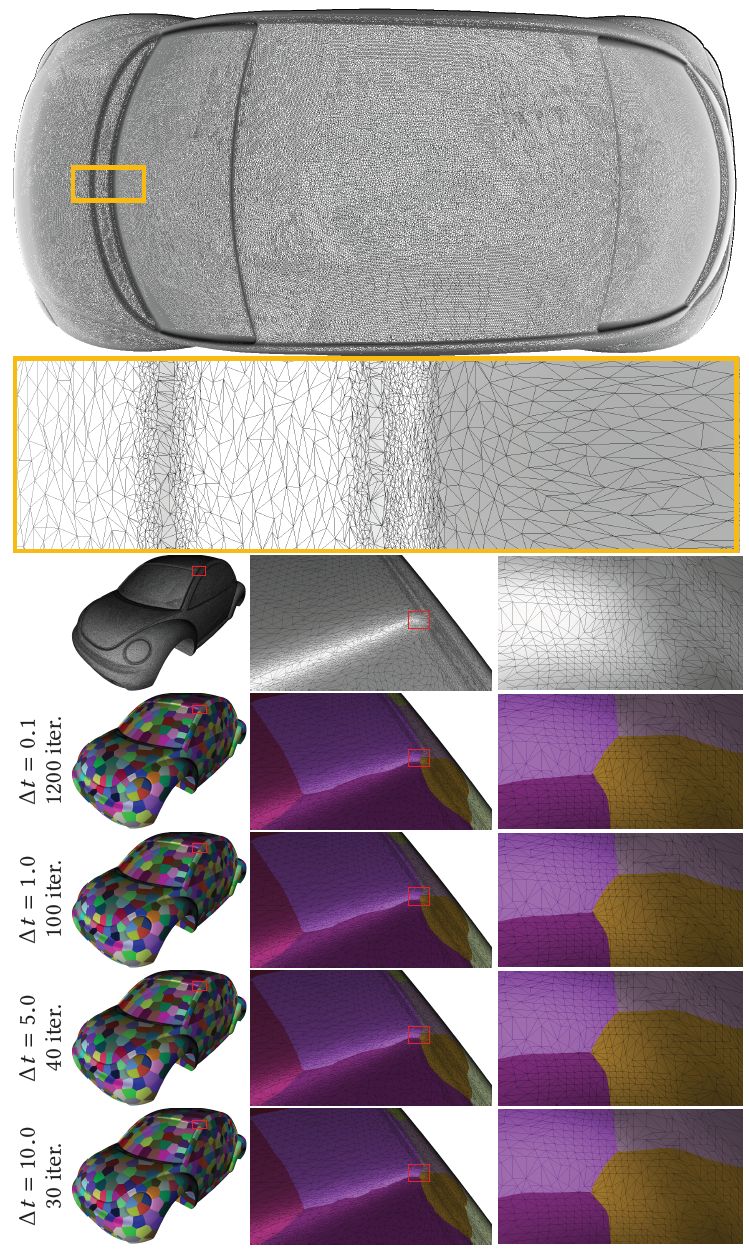}
	
	\caption{(top) The beetle model (2M$\Delta$) and a zoom-in highlighting irregular variation in density and mesh sizing. (bottom) Sharp visualization (field with the largest influence) of the convergence rates for different time steps $0.1$, $1.0$, $5.0$, and $10.0$ for $1200$, $100$, $40$, and $30$ iterations, respectively.
	Even for this highly irregular mesh, there is hardly any difference after convergence for $\Delta t = 0.1 - 5.0$, only at $\Delta t = 10$ differences are noticeable. Please zoom in for details in the electronic version.
	}
	\label{fig:beetle_mesh}
\end{figure}

\section{Convergence and stability analysis}

In our current numerical formulation we adopted a basic Euler step approach. Besides simplicity, this choice is motivated by two main reasons: suitability for the problem at hand and performance gains. Due to its simplicity, it enables our approach to achieve high performance for large meshes sizes---which to the best of our knowledge have not been reached by existing approaches.

The use of an implicit stepping approach such as backward Euler or the Adam-Bashforth approach would require use of iterative or direct solver within each step. This clearly affects performance and limits the size of problems that can be addressed especially if a direct solver is used due to fill-in effect in the factorization. Iterative solvers on the other hand may require a large number of iterations and/or adequate preconditioning. Furthermore, the extension to centroidal Voronoi diagrams where dynamic updates are performed would not be as straightforward as with our approach. In this respect, we would like to emphasize that the approach proposed by \citeauthor{Herholz:dd:2017}~\shortcite{Herholz:dd:2017} is not an implicit variant of ours. In fact, it is not governed by a time dependent equation and their solution is not an implicit approach at all since they solve for each cell separately. Additionally, their cells are not the result of a free evolution but rather by means of an initial estimate of prescribed radii using Dijkstra's shortest path algorithm.

While a theoretical stability analysis for the governing equations of our current system (equations~\ref{equ:phitimeDerivativeMain},~\ref{equ:dfe},~\ref{equ:Eulerstep}) is beyond the scope of this work given the difficulty of such exercise for nonlinear problems, see for instance \citeauthor{leveque_2002} \shortcite{leveque_2002}. Our empirical tests suggest that the problem is not stiff and for varying time steps the resulting tessellations are remarkably identical. A very challenging case is summarized in Figure~\ref{fig:beetle_mesh} for a the beetle model which features a highly irregular triangulation with varying densities.
As can be seen, a variation of the time step across a factor of $50$ achieves very similar results. Only, when increasing the time step further to $\Delta t = 10.0$ visible differences show up when extracting the exact border between cells.
Note that the number of iterations required does not significantly reduce when going from $\Delta t = 5.0$ to $\Delta t = 10.0$, which indicates, that $\Delta t$ is getting closer to the usable maximum, as triangles always require one iteration to propagate the field.
Nevertheless, the results are still remarkably similar.

\begin{figure}
	\centering
	\includegraphics[width=.98\linewidth]{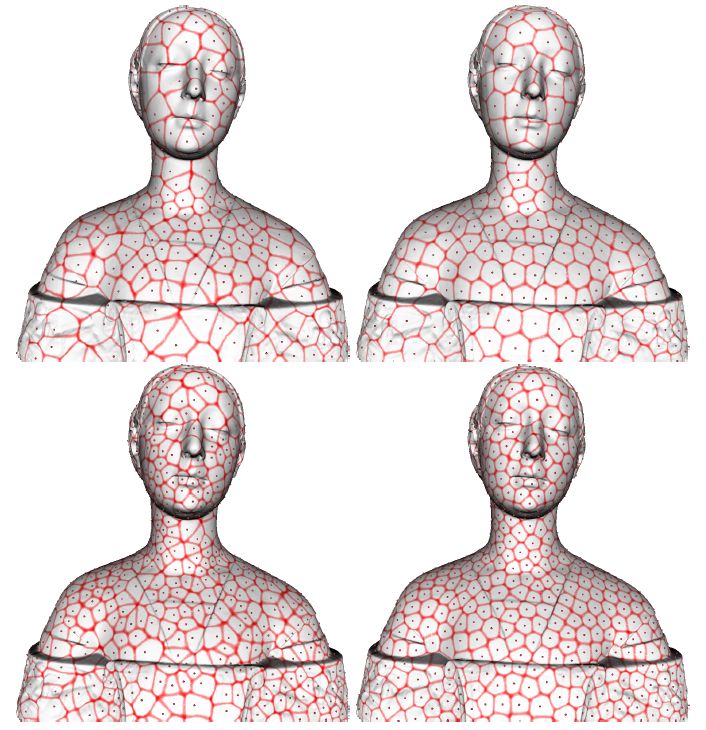}
	\caption{The initial tessellation (left) obtained from an increasing number of seeds ($500, 1000$ from top) randomly distributed over the Sforza model (500K$\Delta$) and the resulting tessellation after 100 Lloyd-like iterations.}
	\label{fig:Sforza}
\end{figure}

\def\rtbs{\hspace{2.4pt}}
\begin{table}
	 \begin{tabular}{@{\rtbs}r@{\rtbs}@{\rtbs}c@{\rtbs}@{\rtbs}r@{\rtbs}@{\rtbs}r@{\rtbs}@{\rtbs}r@{\rtbs}@{\rtbs}r@{\rtbs}@{\rtbs}r@{\rtbs}@{\rtbs}r@{\rtbs}}
		\toprule
		Model & Faces & Seeds & $t_{min}$ & $t_{max}$ & $t_{mean}$ & $t_{conv}$ & Memory \\
		\midrule
		\multirow{2}{*}{Tweety} & \multirow{2}{*}{54k} & 500  & 1.22 & 6.95 & 2.44 & 366 & 49 \\ \vspace{2pt}
		                        &                      & 2000 & 6.21 & 45.03 & 19.99 & 296 & 184 \\
		\multirow{2}{*}{Sforza} & \multirow{2}{*}{500k}& 1000 & 2.45 & 12.59 & 4.56 & 672 & 197 \\ \vspace{2pt}
	                            &                      & 5000 & 2.57 & 30.87 & 12.52 & 1250 & 443 \\
	    \multirow{2}{*}{Hand}   & \multirow{2}{*}{1.5M}& 1000 & 5.05 & 20.27 & 8.46 & 2544 & 247 \\ \vspace{2pt}
	                            &                      & 5000 & 5.67 & 37.81 & 15.20 & 2251 & 705 \\
	    \multirow{2}{*}{Beetle} & \multirow{2}{*}{2.0M}& 1000 & 6.21 & 20.39 & 10.40 & 3132 & 361 \\ \vspace{2pt}
	   						    &                      & 5000 & 6.88 & 38.10 & 18.13 & 2718 & 940 \\
	   	\multirow{2}{*}{Bimba}  & \multirow{2}{*}{3.7M}& 5000 & 10.73 & 51.15 & 24.01 & 4793 & 1091 \\ \vspace{2pt}
	   						    &                      & 7500 & 11.76 & 55.35 & 27.29 & 4122 & 1689 \\
	   	\multirow{2}{*}{Orchid} & \multirow{2}{*}{4.0M}& 5000 & 12.06 & 58.24 & 25.76 & 5129 & 1513 \\ \vspace{2pt}
	   	                        &                      & 7500 & 12.94 & 60.27 & 29.39 & 4443 & 2031 \\
	    \multirow{2}{*}{Flightsuit} & \multirow{2}{*}{21.4M}& 7500 &  62.30 & 139.55 & 106.92 & 4761 & 3542 \\ \vspace{2pt}
	   	                        &                      & 10000 & 64.89 & 157.52 & 125.13 & 3064 & 4838 \\
	    \bottomrule
	\end{tabular}
	\caption{Performance comparison for various test models and seed numbers. Timings are provided in $ms$; $t_{min}$, $t_{max}$, and $t_{mean}$ are the respective step timings for one Euler step; $t_{conv}$ is the average time ($ms$) for one Lloyd-like iteration, \ie, the sum of step timings until the fields are converged. Memory lists the peak memory consumption during the executed steps.}
	\label{tab:restable}
\end{table}

\begin{figure}
	\centering
	 \hspace{12pt}\includegraphics[width=0.6\linewidth]{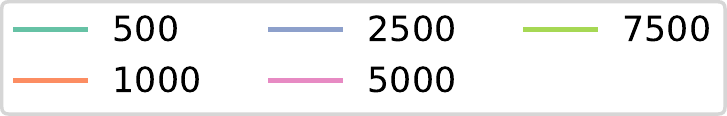}\vspace{-5pt}\\
	\begin{subfigure}{\linewidth}
	\includegraphics[width=\linewidth]{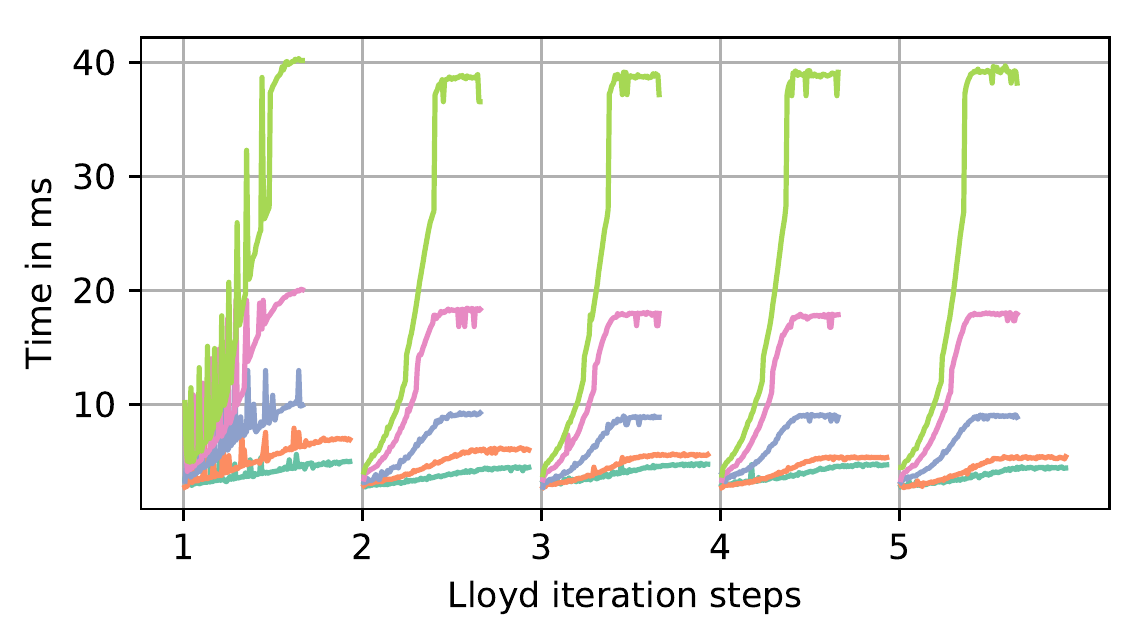}\\
	\vspace{-18pt}
	\caption{Sforza (500K$\Delta$)}
	\end{subfigure}
	\begin{subfigure}{\linewidth}
	\includegraphics[width=\linewidth]{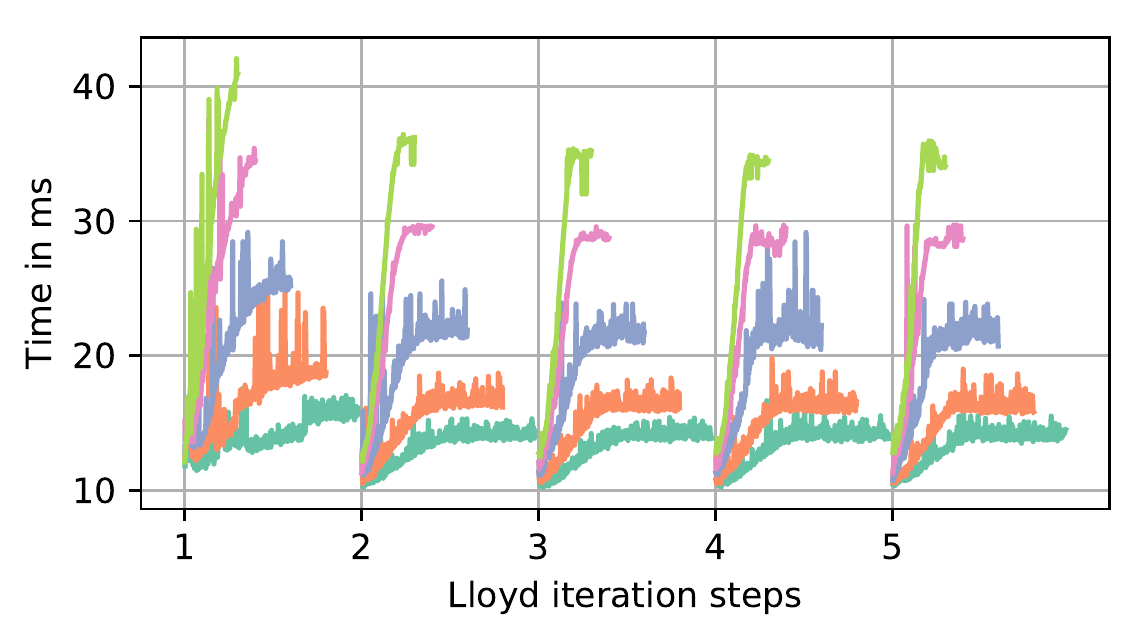}\\
	\vspace{-18pt}
	\caption{Bimba (3.7M$\Delta$)}
	\end{subfigure}\\
	\vspace{-10pt}
	\caption{Detailed step timings in $ms$ for the first five Lloyd iterations with different numbers of seeds ($500,1000,2500,5000,7500$) for the Sforza and Bimba models.}
	\label{fig:timing}
\end{figure}

%

\begin{figure}
	\centering
    \includegraphics[width=.98\linewidth] {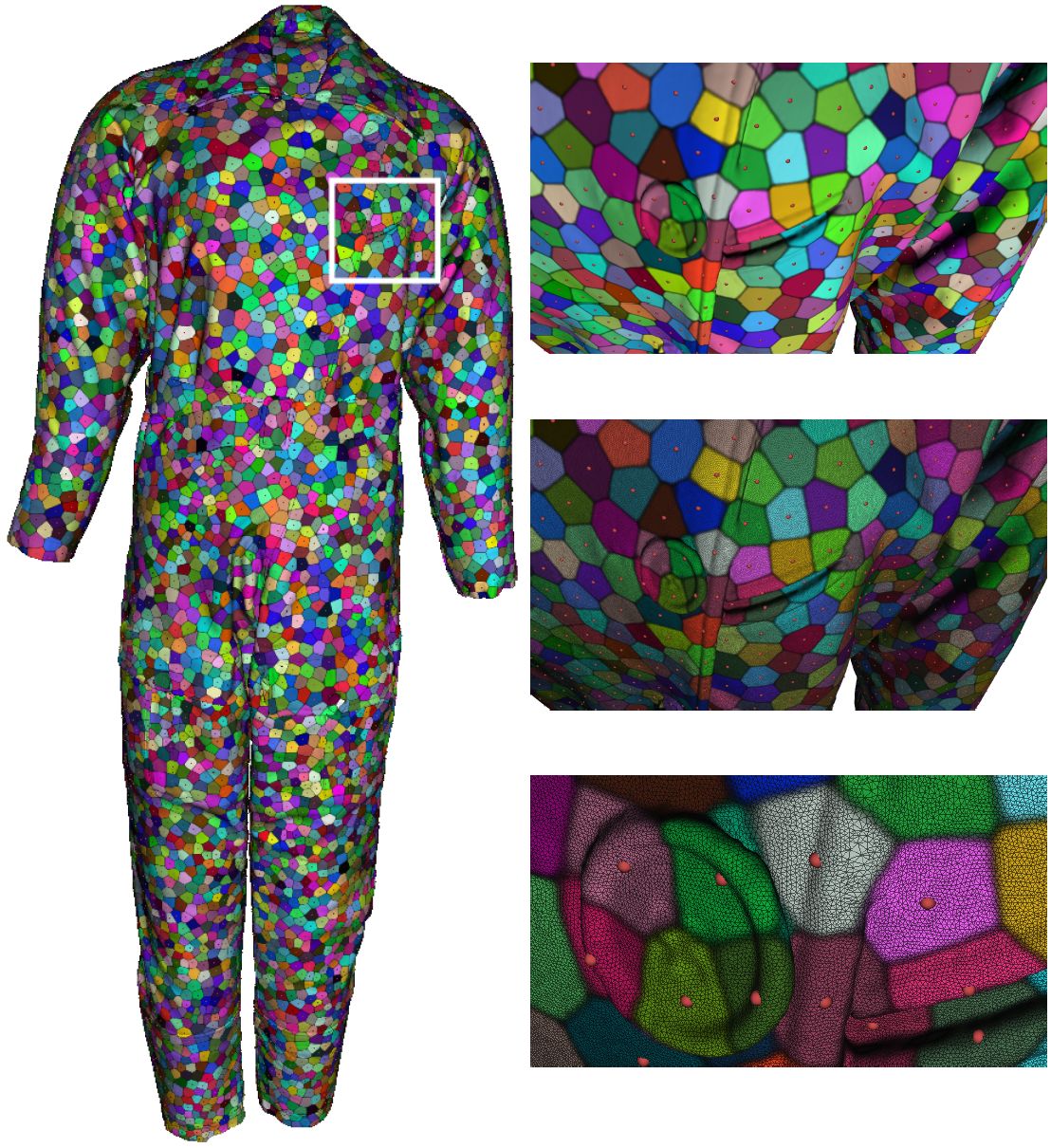}

	\caption{Earhart's Flight Suit (10M$\Delta$) with $10\,000$ seeds, the resulting cells are visualized in different colors. A close-up of the cell distribution (top-row) and the underlying mesh (middle and bottom rows) reveals small scale geometric complexity (pocket fold, button fold) that can successfully be processed with our computationally- and memory efficient approach.}
	\label{fig:flightsuit}
\end{figure}

\section{Results}
\label{sec:results}

Throughout our experiments, we used the following hardware configuration: an Intel i7 6800k CPU with 32GB of memory and an NVIDIA Geforce Titan Xp with 3840 compute cores and 12GB of memory.
We used $\Delta t=5.0$.

\subsection{Field propagation and Lloyd-like iterations}
\begin{figure}
	\centering
	\includegraphics[width=.98\linewidth]{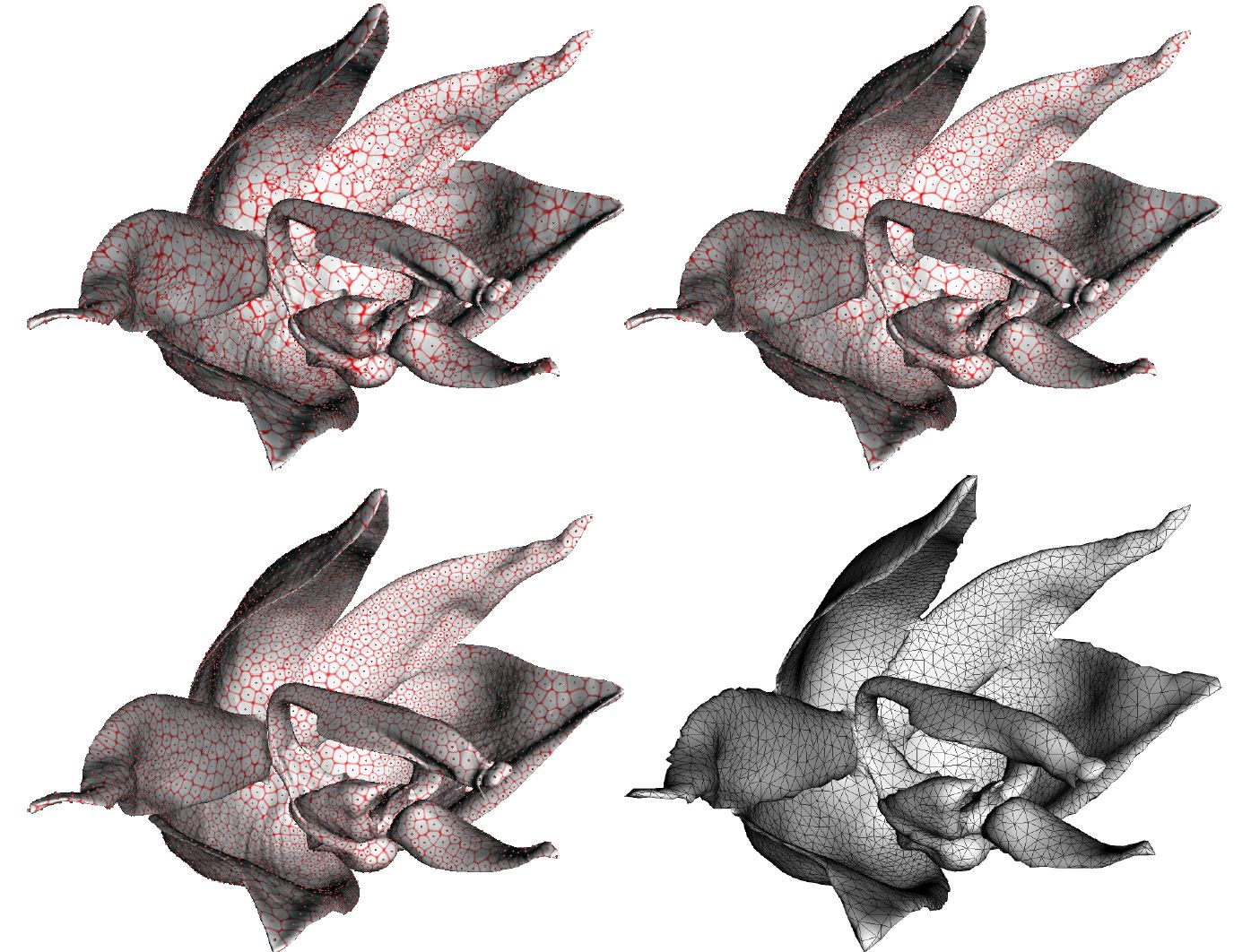}%
	\caption{The progress of the propagation from an initial random distribution of 7500 seeds on the Eembreea Orchid model (4M$\Delta$) and the resulting tessellations after one and 100 Lloyd-like iterations and the extracted dual mesh with 15k$\Delta$.}
	\label{fig:Orchid}
\end{figure}

A performance breakdown of our approach and an overview of the used models is given Table~\ref{tab:restable}. Renderings of the models are found in Figure~\ref{fig:beetle}, \ref{fig:Sforza}, \ref{fig:bimba}, \ref{fig:Orchid}, and \ref{fig:flightsuit}.
Detailed step timings are given in Figure~\ref{fig:timing}.
As can be seen, the per-step timings increase with each iteration until the fields converge.
This fact can be attributed to $\mathbf{\Phi}$ growing in size as the fields propagate, which increases the cost of the sparse matrix-matrix multiplication.
As expected, both, the mesh size as well as the number of seeds influence the step timings.
However, as the number of seeds on a mesh increases, fewer steps are required until the fields converge (also indicated by the graph for higher seeds counts ending early in Figure~\ref{fig:timing}).
Thus, the overall time until field convergence can even reduce with increasing number of seeds.

The spikes in the steps timings indicate reallocation of either of the matrices and thus also an increase in memory consumption.
Nevertheless, our approach can handle very large meshes, like the Amelia Earhart Flight Suit depicted in Figure~\ref{fig:flightsuit}.
The maximum memory consumption is directly related to the mesh size and the number of seeds.

Typical examples of our experimental results are shown in Figures~\ref{fig:beetle}, \ref{fig:Sforza}, and \ref{fig:bimba}. The images show the results for an initial random distribution and the results after a given number of Lloyd-like iterations.
Our approach works on highly detailed models as the Amelia Earhart Flight Suit, as well as very complicated geometric figures such as the Eembreea Orchid model (Figure~\ref{fig:Orchid}).
The Bimba model (Figure~\ref{fig:bimba}) is interesting because it features holes and mesh irregularities in the ear, mouth, and hair regions. Nevertheless our results remain well behaved. Our use of sparse matrices for encoding the layered formulation of our approach offers the flexibility of accessing the individual regions and extracting smooth and sharp boundaries as illustrated in Figure~\ref{fig:beetle_mesh} and \ref{fig:flightsuit}.

The accompanying video shows that our approach can handle complicated geometric figures and large models without any additional requirements. All parameters defined in our work were kept unchanged throughout all experiments. As our fields propagates directly on the surface, we do not require additional snapping or re-projection to remain on the surface as commonly used in other approaches.

\subsection{Dual-mesh extraction}
\begin{figure*}
	\centering
       \includegraphics[width=.98\linewidth] {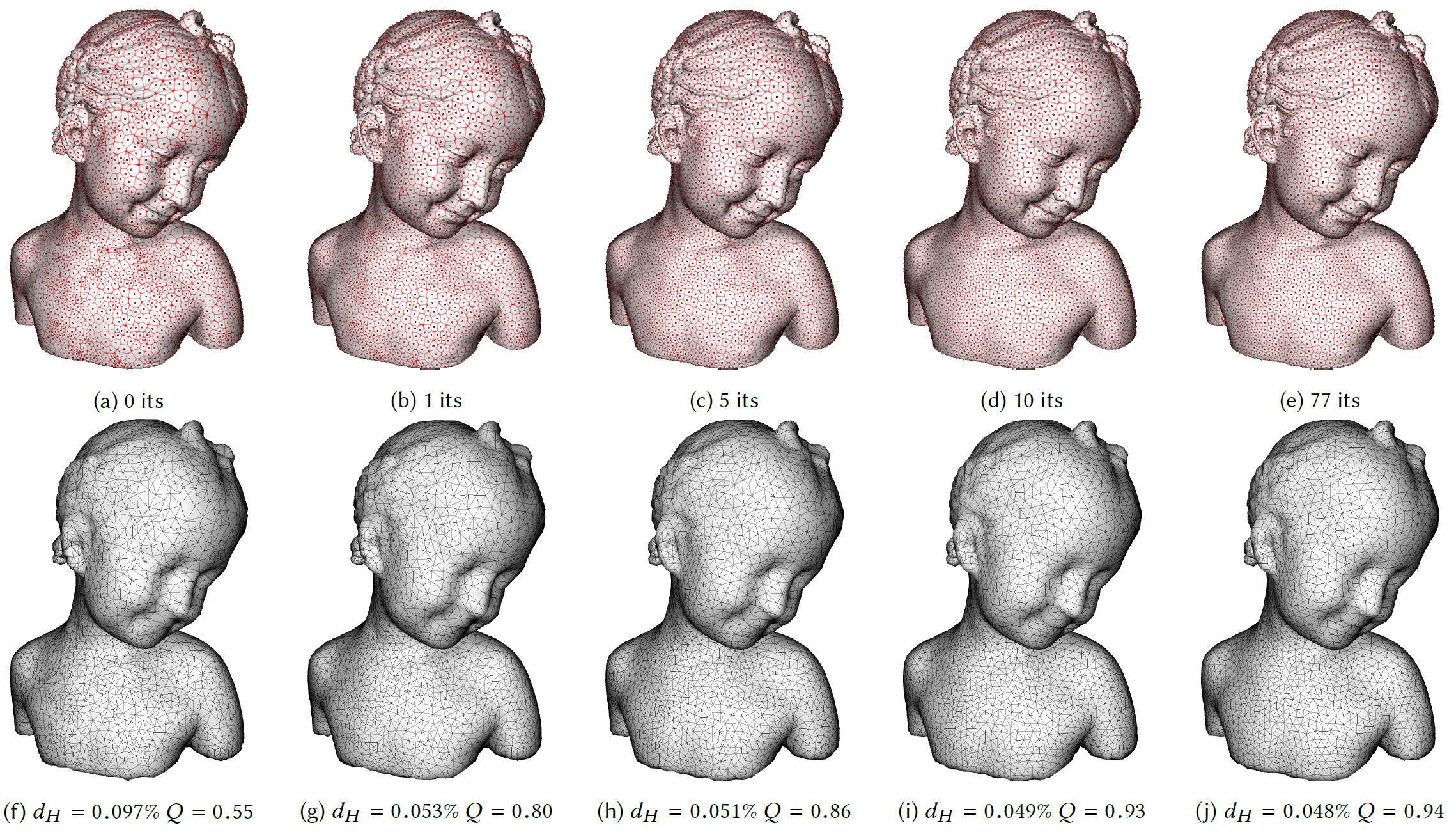}

	\caption{Five selected iterations of our Lloyd-like algorithm on the Bimba model ($3.7M\Delta$) and the extracted meshes. Clearly, the extracted meshes increase in quality and get closer to the original model as the number of iterations increases.}
	\label{fig:bimba}
\end{figure*}

Dual-mesh extraction examples are shown in Figure~\ref{fig:Orchid} and \ref{fig:bimba}. Table~\ref{tab:meshextraction} shows quality metrics computed for the different test cases and various seed constellations.
Successful Lloyd iterations ideally distribute the seeds in such away across the model that the extracted dual meshes match the original model more closely.
At the same time, the mesh quality should also increase.
As can be seen in the figures and in the tables that is the case.
The mean Hausdorff distance between the extracted dual-mesh and its variance decrease with increasing iterations.
Furthermore, the triangle quality (based on the ratio between circumcircle and incircle radius) increases.
This goes hand in hand with the minimum angle increasing, the maximum angle decreasing and the number of angles below $30^\circ$ lowering.
We recorded consistent behavior throughout all tested meshes.

\def\rtbs{\hspace{2.1pt}}
\def\rtbl{\hspace{5.5pt}}
\def\ltbl{\vspace{2.8pt}}

\begin{table*}
	\centering
	 \begin{tabular}{@{\rtbs}l@{\rtbs}@{\rtbs}r@{\rtbs}@{\rtbs}r@{\rtbs}@{\rtbs}r@{\rtbl}@{\rtbl}r@{\rtbs}@{\rtbs}r@{\rtbs}@{\rtbs}r@{\rtbl}@{\rtbl}r@{\rtbs}@{\rtbs}r@{\rtbs}@{\rtbs}r@{\rtbl}@{\rtbl}r@{\rtbs}@{\rtbs}r@{\rtbs}@{\rtbs}r@{\rtbl}@{\rtbl}r@{\rtbs}@{\rtbs}r@{\rtbs}@{\rtbs}r@{\rtbl}@{\rtbl}r@{\rtbs}@{\rtbs}r@{\rtbs}@{\rtbs}r@{\rtbs}}
		\toprule
		& \multicolumn{3}{c}{Tweety 100} & \multicolumn{3}{c}{Sforza 1000} & \multicolumn{3}{c}{Hand 1000} & \multicolumn{3}{c}{Beetle 5000} & \multicolumn{3}{c}{Bimba 5000} & \multicolumn{3}{c}{Orchid 7500} \\
		Lloyd iterations
		& \multicolumn{1}{@{\rtbs}c@{\rtbs}}{0} & \multicolumn{1}{@{\rtbs}c@{\rtbs}}{1} & \multicolumn{1}{@{\rtbs}c@{\rtbs}}{21} & \multicolumn{1}{@{\rtbs}c@{\rtbs}}{0} & \multicolumn{1}{@{\rtbs}c@{\rtbs}}{1} & \multicolumn{1}{@{\rtbs}c@{\rtbs}}{61} & \multicolumn{1}{@{\rtbs}c@{\rtbs}}{0} & \multicolumn{1}{@{\rtbs}c@{\rtbs}}{1} & \multicolumn{1}{@{\rtbs}c@{\rtbs}}{79} & \multicolumn{1}{@{\rtbs}c@{\rtbs}}{0} & \multicolumn{1}{@{\rtbs}c@{\rtbs}}{1} & \multicolumn{1}{@{\rtbs}c@{\rtbs}}{81} & \multicolumn{1}{@{\rtbs}c@{\rtbs}}{0} & \multicolumn{1}{@{\rtbs}c@{\rtbs}}{1} & \multicolumn{1}{@{\rtbs}c@{\rtbs}}{77} & \multicolumn{1}{@{\rtbs}c@{\rtbs}}{0} & \multicolumn{1}{c}{1} & \multicolumn{1}{@{\rtbs}c@{\rtbs}}{93} \\
		\midrule
		Dist (\%) &       1.14 & 0.90  & 0.89  &  0.20 &  0.15 & 0.14    & 0.38 & 0.21 & 0.83    &    0.10 & 0.08 & 0.07    & 0.09 & 0.05 & 0.04 &    0.11 & 0.06 & 0.06\\
		Dist RMS (\%) &   1.42 & 1.19  & 1.19  &  0.35 &  0.26 & 0.23    & 0.63 & 0.37 & 0.26    &    0.19 & 0.14 & 0.11    & 0.15 & 0.09 & 0.08 &    0.20 & 0.11 & 0.11\\
		Mean Quality &    0.62 & 0.87  & 0.91  &  0.80 &  0.86 & 0.93    & 0.82 & 0.87 & 0.96    &    0.77 & 0.84 & 0.93    & 0.81 & 0.87 & 0.94 &    0.64 & 0.79 & 0.89 \\
		Min Quality &     0.11 & 0.37  & 0.43  &  0.01 &  0.13 & 0.41    & 0.03 & 0.20 & 0.63    &    0.01 & 0.31 & 0.62    & 0.02 & 0.06 & 0.34 &    0.01 & 0.05 & 0.26 \\
		Mean Min Angle & 30.62 & 38.87 & 45.85 & 30.16 & 39.89 & 47.16   & 30.24 & 39.69 & 53.45 &   26.87 & 36.61 & 45.64  & 39.24 & 43.17 & 51.11 & 29.16 & 35.38 & 45.32 \\
		Min Angle &        12.3 & 19.8 & 37.0 &   3.8 & 14.5 & 37.4      & 5.3 & 18.5 & 44.1     &     1.3 & 8.6 & 28.8     & 0.9 & 8.5 & 29.2 &      0.4 & 2.4 & 24.9 \\
		$<30^\circ$ (\%)& 44.80 & 4.59 & 0.0 & 52.97 & 5.68 & 0.0      & 54.70 & 4.61 & 0.00   &   60.16 & 12.10 & 1.17   & 19.69 & 5.49 & 0.11 &    56.24 & 24.00 & 5.06 \\
		\bottomrule
	\end{tabular}
	\caption{Quality metrics for the dual meshes extracted from various models for increasing number of Lloyd-like iterations (till convergence). Dist corresponds to the Hausdorff distance between the original and the dual mesh relative to the model's diagonal (lower is better); Quality is the average ratio between circumcircle and incircle radius for all triangles ($1.0$ is best); Mean Min Angle corresponds to the average over the minimum angle in each triangle ($60^\circ$ is best).}
	\vspace{-8pt}
	\label{tab:meshextraction}
\end{table*}

\section{Conclusion}

The overall objective of this work was to show that by changing the way we apprehend natural tessellations on surfaces a different numerical model, which breaks away from the centuries old straight line diagrams, is possible. In this model, tessellations emerge as the solution of a system of time dependent partial differential equations layered on the surface.

The emphasis of this work was on supplying the underlying numerical machinery which allows for representing and manipulating the resulting cells efficiently.
As a proof of concept, we showed throughout a simple extension to LLoyd's-like iterations that intricate geometric and topological operations pertining to cell updates and dual mesh extraction can be translated into simple linear algebra formulations. Despite this simplicity, our solution is comprehensive and has minimal requirements.

Channeling the numerical effort to linear algebra kernels allowed for taking advantage of fine grained parallelism of modern GPUs. In this way, models encompassing millions of faces can be processed at unprecedented rates allowing for direct monitoring of various aspects of the tessellation prrocess.

As a venue of future work, we will explore anisotropic field control on surfaces as well as extensions to the volumetric setting.

\begin{acks}

Images in Figure~\ref{fig:natureImages} are courtesy of Laszlo Ilyes
(Shell Games), Neil Palmer (cracked earth), and Rouven Kurz (Giraffen) under the creative commons license.

\end{acks}

\bibliographystyle{ACM-Reference-Format}
\bibliography{}


\begin{thebibliography}{30}


\ifx \showCODEN    \undefined \def \showCODEN     #1{\unskip}     \fi
\ifx \showDOI      \undefined \def \showDOI       #1{#1}\fi
\ifx \showISBNx    \undefined \def \showISBNx     #1{\unskip}     \fi
\ifx \showISBNxiii \undefined \def \showISBNxiii  #1{\unskip}     \fi
\ifx \showISSN     \undefined \def \showISSN      #1{\unskip}     \fi
\ifx \showLCCN     \undefined \def \showLCCN      #1{\unskip}     \fi
\ifx \shownote     \undefined \def \shownote      #1{#1}          \fi
\ifx \showarticletitle \undefined \def \showarticletitle #1{#1}   \fi
\ifx \showURL      \undefined \def \showURL       {\relax}        \fi
\providecommand\bibfield[2]{#2}
\providecommand\bibinfo[2]{#2}
\providecommand\natexlab[1]{#1}
\providecommand\showeprint[2][]{arXiv:#2}

\bibitem[\protect\citeauthoryear{Alliez, Verdi\`{e}re, Devillers, and
  Isenburg}{Alliez et~al\mbox{.}}{2003}]%
        {Alliez:2003:ISR}
\bibfield{author}{\bibinfo{person}{Pierre Alliez}, \bibinfo{person}{\'{E}ric
  Colin~de Verdi\`{e}re}, \bibinfo{person}{Olivier Devillers}, {and}
  \bibinfo{person}{Martin Isenburg}.} \bibinfo{year}{2003}\natexlab{}.
\newblock \showarticletitle{Isotropic Surface Remeshing}. In
  \bibinfo{booktitle}{{\em Proceedings of the Shape Modeling International
  2003}} {\em (\bibinfo{series}{SMI '03})}. \bibinfo{publisher}{IEEE Computer
  Society}, \bibinfo{address}{Washington, DC, USA}, \bibinfo{pages}{49--58}.
\newblock
\showISBNx{0-7695-1909-1}


\bibitem[\protect\citeauthoryear{Crane, Weischedel, and Wardetzky}{Crane
  et~al\mbox{.}}{2013}]%
        {Crane:2013:GHN}
\bibfield{author}{\bibinfo{person}{Keenan Crane}, \bibinfo{person}{Clarisse
  Weischedel}, {and} \bibinfo{person}{Max Wardetzky}.}
  \bibinfo{year}{2013}\natexlab{}.
\newblock \showarticletitle{Geodesics in Heat: A New Approach to Computing
  Distance Based on Heat Flow}.
\newblock \bibinfo{journal}{{\em ACM Trans. Graph.\/}} \bibinfo{volume}{32},
  \bibinfo{number}{5}, Article \bibinfo{articleno}{152} (\bibinfo{date}{Oct.}
  \bibinfo{year}{2013}), \bibinfo{numpages}{11}~pages.
\newblock
\showISSN{0730-0301}


\bibitem[\protect\citeauthoryear{Du and Emelianenko}{Du and
  Emelianenko}{2006}]%
        {Du:2006:ASCCVT}
\bibfield{author}{\bibinfo{person}{Qiang Du} {and} \bibinfo{person}{Maria
  Emelianenko}.} \bibinfo{year}{2006}\natexlab{}.
\newblock \showarticletitle{Acceleration schemes for computing centroidal
  Voronoi tessellations}.
\newblock \bibinfo{journal}{{\em Numerical Linear Algebra with Applications\/}}
  \bibinfo{volume}{13}, \bibinfo{number}{2-3} (\bibinfo{year}{2006}),
  \bibinfo{pages}{173--192}.
\newblock
\showISSN{1099-1506}
\showDOI{%
\url{https://doi.org/10.1002/nla.476}}


\bibitem[\protect\citeauthoryear{Du, Gunzburger, and Ju}{Du
  et~al\mbox{.}}{2003}]%
        {Du:2003:CCVTS}
\bibfield{author}{\bibinfo{person}{Qiang Du}, \bibinfo{person}{Max~D.
  Gunzburger}, {and} \bibinfo{person}{Lili Ju}.}
  \bibinfo{year}{2003}\natexlab{}.
\newblock \showarticletitle{Constrained Centroidal Voronoi Tessellations for
  Surfaces}.
\newblock \bibinfo{journal}{{\em SIAM Journal on Scientific Computing\/}}
  \bibinfo{volume}{24}, \bibinfo{number}{5} (\bibinfo{year}{2003}),
  \bibinfo{pages}{1488--1506}.
\newblock
\showDOI{%
\url{https://doi.org/10.1137/S1064827501391576}}


\bibitem[\protect\citeauthoryear{Fix}{Fix}{1981}]%
        {Fix:1981:PFM}
\bibfield{author}{\bibinfo{person}{G.~J. Fix}.}
  \bibinfo{year}{1981}\natexlab{}.
\newblock \showarticletitle{Phase field methods for free boundary problems}. In
  \bibinfo{booktitle}{{\em Proceedings of Interdisciplinary Symposium on Free
  Boundary Problems: Theory and Applications}} {\em (\bibinfo{series}{Research
  Notes in Mathematics})}, \bibfield{editor}{\bibinfo{person}{(A. Fasano} {and}
  \bibinfo{person}{M.~Primicerio}} (Eds.), Vol.~\bibinfo{volume}{78, 79}.
  \bibinfo{pages}{580--589}.
\newblock


\bibitem[\protect\citeauthoryear{Gersho}{Gersho}{1979}]%
        {Gersho:1979:AOBQ}
\bibfield{author}{\bibinfo{person}{A. Gersho}.}
  \bibinfo{year}{1979}\natexlab{}.
\newblock \showarticletitle{Asymptotically optimal block quantization}.
\newblock \bibinfo{journal}{{\em IEEE Transactions on Information Theory\/}}
  \bibinfo{volume}{25}, \bibinfo{number}{4} (\bibinfo{date}{Jul}
  \bibinfo{year}{1979}), \bibinfo{pages}{373--380}.
\newblock
\showISSN{0018-9448}


\bibitem[\protect\citeauthoryear{Herholz, Haase, and Alexa}{Herholz
  et~al\mbox{.}}{2017}]%
        {Herholz:dd:2017}
\bibfield{author}{\bibinfo{person}{Philipp Herholz}, \bibinfo{person}{Felix
  Haase}, {and} \bibinfo{person}{Marc Alexa}.} \bibinfo{year}{2017}\natexlab{}.
\newblock \showarticletitle{{Diffusion Diagrams: Voronoi Cells and Centroids
  from Diffusion}}.
\newblock \bibinfo{journal}{{\em Computer Graphics Forum\/}}
  (\bibinfo{year}{2017}).
\newblock
\showISSN{1467-8659}
\showDOI{%
\url{https://doi.org/10.1111/cgf.13116}}


\bibitem[\protect\citeauthoryear{Iri, Murota, and Ohya}{Iri
  et~al\mbox{.}}{1984}]%
        {Iri:1984}
\bibfield{author}{\bibinfo{person}{Masao Iri}, \bibinfo{person}{Kazuo Murota},
  {and} \bibinfo{person}{Takao Ohya}.} \bibinfo{year}{1984}\natexlab{}.
\newblock \bibinfo{booktitle}{{\em A fast Voronoi-diagram algorithm with
  applications to geographical optimization problems}}.
\newblock \bibinfo{publisher}{Springer Berlin Heidelberg},
  \bibinfo{address}{Berlin, Heidelberg}, \bibinfo{pages}{273--288}.
\newblock
\showISBNx{978-3-540-38828-9}


\bibitem[\protect\citeauthoryear{Ju, Du, and Gunzburger}{Ju
  et~al\mbox{.}}{2002}]%
        {Ju:2002:PMCVT}
\bibfield{author}{\bibinfo{person}{Lili Ju}, \bibinfo{person}{Qiang Du}, {and}
  \bibinfo{person}{Max Gunzburger}.} \bibinfo{year}{2002}\natexlab{}.
\newblock \showarticletitle{Probabilistic methods for centroidal Voronoi
  tessellations and their parallel implementations}.
\newblock \bibinfo{journal}{{\it Parallel Comput.}} \bibinfo{volume}{28},
  \bibinfo{number}{10} (\bibinfo{year}{2002}), \bibinfo{pages}{1477 -- 1500}.
\newblock
\showISSN{0167-8191}


\bibitem[\protect\citeauthoryear{Kimmel and Sethian}{Kimmel and
  Sethian}{1998}]%
        {Kimmel:98:CGPM}
\bibfield{author}{\bibinfo{person}{R. Kimmel} {and} \bibinfo{person}{J.~A.
  Sethian}.} \bibinfo{year}{1998}\natexlab{}.
\newblock \showarticletitle{Computing Geodesic Paths on Manifolds}. In
  \bibinfo{booktitle}{{\em Proc. Natl. Acad. Sci. USA}}.
  \bibinfo{pages}{8431--8435}.
\newblock


\bibitem[\protect\citeauthoryear{LeVeque}{LeVeque}{2002}]%
        {leveque_2002}
\bibfield{author}{\bibinfo{person}{Randall~J. LeVeque}.}
  \bibinfo{year}{2002}\natexlab{}.
\newblock \bibinfo{booktitle}{{\em Finite Volume Methods for Hyperbolic
  Problems}}.
\newblock \bibinfo{publisher}{Cambridge University Press}.
\newblock
\showDOI{%
\url{https://doi.org/10.1017/CBO9780511791253}}


\bibitem[\protect\citeauthoryear{Liu, Wang, L{\'e}vy, Sun, Yan, Lu, and
  Yang}{Liu et~al\mbox{.}}{2009}]%
        {Liu:2009:CVT}
\bibfield{author}{\bibinfo{person}{Yang Liu}, \bibinfo{person}{Wenping Wang},
  \bibinfo{person}{Bruno L{\'e}vy}, \bibinfo{person}{Feng Sun},
  \bibinfo{person}{Dong-Ming Yan}, \bibinfo{person}{Lin Lu}, {and}
  \bibinfo{person}{Chenglei Yang}.} \bibinfo{year}{2009}\natexlab{}.
\newblock \showarticletitle{On Centroidal Voronoi Tessellation\&Mdash;Energy
  Smoothness and Fast Computation}.
\newblock \bibinfo{journal}{{\em ACM Trans. Graph.\/}} \bibinfo{volume}{28},
  \bibinfo{number}{4}, Article \bibinfo{articleno}{101} (\bibinfo{date}{Sept.}
  \bibinfo{year}{2009}), \bibinfo{numpages}{17}~pages.
\newblock
\showISSN{0730-0301}
\showDOI{%
\url{https://doi.org/10.1145/1559755.1559758}}


\bibitem[\protect\citeauthoryear{Lloyd}{Lloyd}{1982}]%
        {Lloyd:1982:LSQ}
\bibfield{author}{\bibinfo{person}{S. Lloyd}.} \bibinfo{year}{1982}\natexlab{}.
\newblock \showarticletitle{Least squares quantization in PCM}.
\newblock \bibinfo{journal}{{\em IEEE Transactions on Information Theory\/}}
  \bibinfo{volume}{28}, \bibinfo{number}{2} (\bibinfo{date}{Mar}
  \bibinfo{year}{1982}), \bibinfo{pages}{129--137}.
\newblock
\showISSN{0018-9448}


\bibitem[\protect\citeauthoryear{MacQueen}{MacQueen}{1967}]%
        {macqueen1967}
\bibfield{author}{\bibinfo{person}{J. MacQueen}.}
  \bibinfo{year}{1967}\natexlab{}.
\newblock \showarticletitle{Some methods for classification and analysis of
  multivariate observations}. In \bibinfo{booktitle}{{\em Proceedings of the
  Fifth Berkeley Symposium on Mathematical Statistics and Probability, Volume
  1: Statistics}}. \bibinfo{publisher}{University of California Press},
  \bibinfo{address}{Berkeley, Calif.}, \bibinfo{pages}{281--297}.
\newblock


\bibitem[\protect\citeauthoryear{NVIDIA}{NVIDIA}{2016}]%
        {Nvidia:cusparse}
\bibfield{author}{\bibinfo{person}{NVIDIA}.} \bibinfo{year}{2016}\natexlab{}.
\newblock \bibinfo{booktitle}{{\em The API reference guide for cuSPARSE, the
  CUDA sparse matrix library.\/} (\bibinfo{edition}{v8.0} ed.)}.
\newblock NVIDIA.
\newblock


\bibitem[\protect\citeauthoryear{Okabe, Boots, and Sugihara}{Okabe
  et~al\mbox{.}}{1992}]%
        {Okabe:1992:STC}
\bibfield{author}{\bibinfo{person}{Atsuyuki Okabe}, \bibinfo{person}{Barry
  Boots}, {and} \bibinfo{person}{Kokichi Sugihara}.}
  \bibinfo{year}{1992}\natexlab{}.
\newblock \bibinfo{booktitle}{{\em Spatial Tessellations: Concepts and
  Applications of Voronoi Diagrams}}.
\newblock \bibinfo{publisher}{John Wiley \& Sons, Inc.}, \bibinfo{address}{New
  York, NY, USA}.
\newblock
\showISBNx{0-471-93430-5}


\bibitem[\protect\citeauthoryear{Peyr{\'e} and Cohen}{Peyr{\'e} and
  Cohen}{2006}]%
        {Peyre:2006}
\bibfield{author}{\bibinfo{person}{Gabriel Peyr{\'e}} {and}
  \bibinfo{person}{Laurent~D. Cohen}.} \bibinfo{year}{2006}\natexlab{}.
\newblock \showarticletitle{Geodesic Remeshing Using Front Propagation}.
\newblock \bibinfo{journal}{{\em International Journal of Computer Vision\/}}
  \bibinfo{volume}{69}, \bibinfo{number}{1} (\bibinfo{year}{2006}),
  \bibinfo{pages}{145}.
\newblock
\showISSN{1573-1405}


\bibitem[\protect\citeauthoryear{Renka}{Renka}{1997}]%
        {Renka:1997:ASD}
\bibfield{author}{\bibinfo{person}{Robert~J. Renka}.}
  \bibinfo{year}{1997}\natexlab{}.
\newblock \showarticletitle{Algorithm 772: STRIPACK: Delaunay Triangulation and
  Voronoi Diagram on the Surface of a Sphere}.
\newblock \bibinfo{journal}{{\em ACM Trans. Math. Softw.\/}}
  \bibinfo{volume}{23}, \bibinfo{number}{3} (\bibinfo{date}{Sept.}
  \bibinfo{year}{1997}), \bibinfo{pages}{416--434}.
\newblock
\showISSN{0098-3500}
\showDOI{%
\url{https://doi.org/10.1145/275323.275329}}


\bibitem[\protect\citeauthoryear{Schall, Zayer, and Seidel}{Schall
  et~al\mbox{.}}{2008}]%
        {Schall08}
\bibfield{author}{\bibinfo{person}{Oliver Schall}, \bibinfo{person}{Rhaleb
  Zayer}, {and} \bibinfo{person}{Hans{-}Peter Seidel}.}
  \bibinfo{year}{2008}\natexlab{}.
\newblock \showarticletitle{Controlled field generation for quad-remeshing}. In
  \bibinfo{booktitle}{{\em Proceedings of the 2008 {ACM} Symposium on Solid and
  Physical Modeling, Stony Brook, New York, USA, June 2-4, 2008}}.
  \bibinfo{pages}{295--300}.
\newblock


\bibitem[\protect\citeauthoryear{Sengupta, Harris, Zhang, and Owens}{Sengupta
  et~al\mbox{.}}{2007}]%
        {Sengupta:2007:SPG}
\bibfield{author}{\bibinfo{person}{S. Sengupta}, \bibinfo{person}{M. Harris},
  \bibinfo{person}{Y. Zhang}, {and} \bibinfo{person}{J.~D. Owens}.}
  \bibinfo{year}{2007}\natexlab{}.
\newblock \showarticletitle{Scan Primitives for GPU Computing}. In
  \bibinfo{booktitle}{{\em Proc. GH '07}}. \bibinfo{pages}{97--106}.
\newblock
\showISBNx{978-1-59593-625-7}


\bibitem[\protect\citeauthoryear{Sethian}{Sethian}{1996}]%
        {Sethian:1996}
\bibfield{author}{\bibinfo{person}{J~A Sethian}.}
  \bibinfo{year}{1996}\natexlab{}.
\newblock \showarticletitle{A fast marching level set method for monotonically
  advancing fronts}.
\newblock \bibinfo{journal}{{\em Proceedings of the National Academy of
  Sciences\/}} \bibinfo{volume}{93}, \bibinfo{number}{4}
  (\bibinfo{year}{1996}), \bibinfo{pages}{1591--1595}.
\newblock


\bibitem[\protect\citeauthoryear{Surazhsky, Alliez, and Gotsman}{Surazhsky
  et~al\mbox{.}}{2003}]%
        {SurazhskyAlliezGotsman2003}
\bibfield{author}{\bibinfo{person}{Vitaly Surazhsky}, \bibinfo{person}{Pierre
  Alliez}, {and} \bibinfo{person}{Craig Gotsman}.}
  \bibinfo{year}{2003}\natexlab{}.
\newblock \showarticletitle{Isotropic remeshing of surfaces: a local
  parameterization approach}. In \bibinfo{booktitle}{{\em Proceedings, 12th
  International Meshing Roundtable}}. \bibinfo{pages}{215--224}.
\newblock


\bibitem[\protect\citeauthoryear{Surazhsky, Surazhsky, Kirsanov, Gortler, and
  Hoppe}{Surazhsky et~al\mbox{.}}{2005}]%
        {Surazhsky:2005:FEAGM}
\bibfield{author}{\bibinfo{person}{Vitaly Surazhsky}, \bibinfo{person}{Tatiana
  Surazhsky}, \bibinfo{person}{Danil Kirsanov}, \bibinfo{person}{Steven~J.
  Gortler}, {and} \bibinfo{person}{Hugues Hoppe}.}
  \bibinfo{year}{2005}\natexlab{}.
\newblock \showarticletitle{Fast Exact and Approximate Geodesics on Meshes}.
\newblock \bibinfo{journal}{{\em ACM Trans. Graph.\/}} \bibinfo{volume}{24},
  \bibinfo{number}{3} (\bibinfo{date}{July} \bibinfo{year}{2005}),
  \bibinfo{pages}{553--560}.
\newblock
\showISSN{0730-0301}
\showDOI{%
\url{https://doi.org/10.1145/1073204.1073228}}


\bibitem[\protect\citeauthoryear{Turk}{Turk}{1991}]%
        {Turk:1991:GTA}
\bibfield{author}{\bibinfo{person}{Greg Turk}.}
  \bibinfo{year}{1991}\natexlab{}.
\newblock \showarticletitle{Generating Textures on Arbitrary Surfaces Using
  Reaction-diffusion}. In \bibinfo{booktitle}{{\em Proceedings of the 18th
  Annual Conference on Computer Graphics and Interactive Techniques}} {\em
  (\bibinfo{series}{SIGGRAPH '91})}. \bibinfo{publisher}{ACM},
  \bibinfo{address}{New York, NY, USA}, \bibinfo{pages}{289--298}.
\newblock
\showISBNx{0-89791-436-8}
\showDOI{%
\url{https://doi.org/10.1145/122718.122749}}


\bibitem[\protect\citeauthoryear{Turk}{Turk}{1992}]%
        {Turk:1992:RPS}
\bibfield{author}{\bibinfo{person}{Greg Turk}.}
  \bibinfo{year}{1992}\natexlab{}.
\newblock \showarticletitle{Re-tiling Polygonal Surfaces}. In
  \bibinfo{booktitle}{{\em Proceedings of the 19th Annual Conference on
  Computer Graphics and Interactive Techniques}} {\em
  (\bibinfo{series}{SIGGRAPH '92})}. \bibinfo{publisher}{ACM},
  \bibinfo{address}{New York, NY, USA}, \bibinfo{pages}{55--64}.
\newblock
\showISBNx{0-89791-479-1}
\showDOI{%
\url{https://doi.org/10.1145/133994.134008}}


\bibitem[\protect\citeauthoryear{Wang, Ying, Liu, Xin, Wang, Gu,
  Mueller-Wittig, and He}{Wang et~al\mbox{.}}{2015}]%
        {Wang:2015:ICCVT}
\bibfield{author}{\bibinfo{person}{Xiaoning Wang}, \bibinfo{person}{Xiang
  Ying}, \bibinfo{person}{Yong-Jin Liu}, \bibinfo{person}{Shi-Qing Xin},
  \bibinfo{person}{Wenping Wang}, \bibinfo{person}{Xianfeng Gu},
  \bibinfo{person}{Wolfgang Mueller-Wittig}, {and} \bibinfo{person}{Ying He}.}
  \bibinfo{year}{2015}\natexlab{}.
\newblock \showarticletitle{Intrinsic computation of centroidal Voronoi
  tessellation (CVT) on meshes}.
\newblock \bibinfo{journal}{{\em Computer-Aided Design\/}}
  \bibinfo{volume}{58} (\bibinfo{year}{2015}), \bibinfo{pages}{51 -- 61}.
\newblock
\showISSN{0010-4485}
\showDOI{%
\url{https://doi.org/10.1016/j.cad.2014.08.023}}
\newblock
\shownote{Solid and Physical Modeling 2014.}


\bibitem[\protect\citeauthoryear{Witkin and Heckbert}{Witkin and
  Heckbert}{1994}]%
        {Witkin:1994}
\bibfield{author}{\bibinfo{person}{Andrew~P. Witkin} {and}
  \bibinfo{person}{Paul~S. Heckbert}.} \bibinfo{year}{1994}\natexlab{}.
\newblock \showarticletitle{Using Particles to Sample and Control Implicit
  Surfaces}. In \bibinfo{booktitle}{{\em Proceedings of the 21st Annual
  Conference on Computer Graphics and Interactive Techniques}} {\em
  (\bibinfo{series}{SIGGRAPH '94})}. \bibinfo{publisher}{ACM},
  \bibinfo{address}{New York, NY, USA}, \bibinfo{pages}{269--277}.
\newblock
\showISBNx{0-89791-667-0}


\bibitem[\protect\citeauthoryear{Xin, L{\'e}vy, Chen, Chu, Yu, Tu, and
  Wang}{Xin et~al\mbox{.}}{2016}]%
        {Xin:2016:CPD}
\bibfield{author}{\bibinfo{person}{Shi-Qing Xin}, \bibinfo{person}{Bruno
  L{\'e}vy}, \bibinfo{person}{Zhonggui Chen}, \bibinfo{person}{Lei Chu},
  \bibinfo{person}{Yaohui Yu}, \bibinfo{person}{Changhe Tu}, {and}
  \bibinfo{person}{Wenping Wang}.} \bibinfo{year}{2016}\natexlab{}.
\newblock \showarticletitle{Centroidal Power Diagrams with Capacity
  Constraints: Computation, Applications, and Extension}.
\newblock \bibinfo{journal}{{\em ACM Trans. Graph.\/}} \bibinfo{volume}{35},
  \bibinfo{number}{6} (\bibinfo{date}{Nov.} \bibinfo{year}{2016}),
  \bibinfo{pages}{244:1--244:12}.
\newblock
\showISSN{0730-0301}


\bibitem[\protect\citeauthoryear{Ying, Wang, and He}{Ying
  et~al\mbox{.}}{2013}]%
        {Ying:2013:SVG}
\bibfield{author}{\bibinfo{person}{Xiang Ying}, \bibinfo{person}{Xiaoning
  Wang}, {and} \bibinfo{person}{Ying He}.} \bibinfo{year}{2013}\natexlab{}.
\newblock \showarticletitle{Saddle Vertex Graph (SVG): A Novel Solution to the
  Discrete Geodesic Problem}.
\newblock \bibinfo{journal}{{\em ACM Trans. Graph.\/}} \bibinfo{volume}{32},
  \bibinfo{number}{6}, Article \bibinfo{articleno}{170} (\bibinfo{date}{Nov.}
  \bibinfo{year}{2013}), \bibinfo{numpages}{12}~pages.
\newblock
\showISSN{0730-0301}


\bibitem[\protect\citeauthoryear{Zayer, Steinberger, and Seidel}{Zayer
  et~al\mbox{.}}{2017}]%
        {meshmatrix17}
\bibfield{author}{\bibinfo{person}{Rhaleb Zayer}, \bibinfo{person}{Markus
  Steinberger}, {and} \bibinfo{person}{Hans-Peter Seidel}.}
  \bibinfo{year}{2017}\natexlab{}.
\newblock \showarticletitle{A GPU-Adapted Structure for Unstructured Grids}.
\newblock \bibinfo{journal}{{\em Computer Graphics Forum\/}}
  \bibinfo{volume}{36}, \bibinfo{number}{2} (\bibinfo{year}{2017}),
  \bibinfo{pages}{495--507}.
\newblock
\showISSN{1467-8659}
\showDOI{%
\url{https://doi.org/10.1111/cgf.13144}}


\end{thebibliography}

\end{document}